\documentclass{aa}  

\usepackage{graphicx}
\usepackage{txfonts}
\usepackage{xcolor}
\usepackage[]{changes}
\definechangesauthor[name=Dominique,color=orange]{dom}
\definechangesauthor[name=Emilie]{emi}

\newcommand{\cmfast}{{\texttt{21cmFAST}}}
\newcommand{\disperse}{{\texttt{DisPerSE}}}
\newcommand{\emma}{{\texttt{EMMA}}}

\usepackage{enumitem}
\begin{document}

   \title{A first look at the topology of reionization redshifts in models of the Epoch of the Reionization}

   \author{Emilie Thélie
          %\inst{1}
          \and
          Dominique Aubert%\inst{1}
          \and
          Nicolas Gillet
          \and
          Pierre Ocvirk
          }

   \institute{Université de Strasbourg, CNRS UMR 7550, Observatoire Astronomique de Strasbourg, Strasbourg, France\\
              \email{emilie.thelie@astro.unistra.fr}}

   \date{Received ..., accepted ...}

% \abstract{}{}{}{}{} 
% 5 {} token are mandatory
 
  \abstract
  % context heading (optional)
  % {} leave it empty if necessary  
   {During the Epoch of Reionization (EoR), the first stars and galaxies appear while creating ionized bubbles that will eventually percolate near $z\sim6$. These ionized bubbles and percolation process are nowadays under a lot of scrutiny since observations of the neutral hydrogen gas will be carried on in the next decade with for example the Square Kilometre Array (SKA) radio-telescope. In the meantime, studies of the EoR are performed on semi-analytical and fully numerical cosmological simulations to investigate for instance the topology of the process. }
  % aims heading (mandatory)
   {We analyse the topology of EoR models through the study of regions that are under the radiative influence of ionization sources. Those regions are associated with peak patches of the reionization redshift field, for which we measure their general properties such as their number, size, shape and orientation. We aim at gaining insights on the geometry of the reionization process and how it relates to the matter distribution for example. We also assess how such measurements can be used to quantify the influence of physical parameters on the reionization models or the differences between fully numerical simulations and semi-analytical models.}
  % methods heading (mandatory)
   {We use the framework of the Morse theory and the persistent homology in the context of the Reionization, which is investigated via the $\disperse$ algorithm on gas density and redshift of reionization maps. We analyse different Reionization scenarios with semi-analytical $\cmfast$ and fully numerical $\emma$ simulations.}
  % results heading (mandatory)
   {We find that we can distinguish between Reionization models with different sources using simple analyses on the number, shape and size distributions of the reionization redshift patches. Meanwhile, for every model (of semi-analytical and fully numerical simulations), we statistically show that those bubbles are rather prolate and aligned with the underlying filaments of gas. Moreover, we briefly highlight that the percolation process of HII bubbles during the EoR can be followed by studying the reionization redshift fields with different persistence thresholds. Finally, we show that fully numerical $\emma$ simulations can be made consistent with $\cmfast$ models in this topological framework as long as the source distribution is diffuse enough.}
  % conclusions heading (optional), leave it empty if necessary 
  {}


   \maketitle
%
%--------------------------------------------------------------------
%--------------------------------------------------------------------
%--------------------------------------------------------------------
\section{Introduction}

The Epoch of Reionization (EoR) marks the transition of a totally neutral to a fully ionized Universe, with its large scales structures composed of filaments, voids and very dense regions. During the EoR, the first sources of radiation appear, releasing the photons required to ionize the cosmic gas. Light escaping from those first stars and galaxies creates HII regions (also known as HII "bubbles") that eventually percolate at the end of Reionization between $z=5.3-6$ \citep{Kulkarni2019}. 

This Epoch is nowadays under many studies to try to understand for instance which kind of sources drove the EoR or how and when the HII bubbles overlap. Analysing the large scale topology of the EoR is useful to unravel the properties of the object that drove the Reionization, such as the nature of the driving sources of the EoR, their spatial distribution, the timing of their appearance or the spectrum of their radiation (see for instance \citet{Chardin2017}). 

Complementary to studies using the power spectrum for instance (e.g. \citet{Mellema2006, Dixon2016, Shaw2020}), many works try to tackle those questions with a topological approach of the epoch of Reionization. For example, Minkowski functionals (and the derived genus or Euler characteristic) are used to analyse the volume, surface aera and curvature of ionized and neutral regions, giving information on the shape of these regions and the percolation process (as in e.g. \citet{Gleser2006, Lee2008, Friedrich2011, Hong2014, Yoshiura2017, Chen2019}). 
Other studies focus on the size distributions of neutral or ionized regions (also respectively called Island Size Distribution or ISD and Bubble Size Distribution or BSD) which characterise the size of neutral islands or HII bubbles and their percolation during the EoR. Examples of methods used in such studies are the Friend-Of-Friend algorithm \citep{Iliev2006, Friedrich2011, Lin2016, Giri2018, Giri2019a}, the Spherical Average Method \citep{Zahn2007, Friedrich2011, Lin2016, Giri2018}, the Mean-Free-Path method \citep{Mesinger2007, Lin2016, Giri2018, Giri2019a}, and the granulometry method \citep{Kakiichi2017}. 
The Betti numbers, related to 3D structures of a field (isolated objects, tunnels and cavities), are also used to probe the topology of ionized bubbles and neutral islands during Reionization \citep{Giri2020}. 
The Triangle Correlation Function (TCF) is another tool to extract topological information, such as the radius and the shape of ionized and neutral bubbles \citep{Gorce2019}. It relies on a 3-points correlation function based on the inverse Fourier transform of the bispectrum that describes the phase of the signal of interest.

Upcoming observations from radio interferometers like the New Extension in Nançay Upgrading loFAR\footnote{https://nenufar.obs-nancay.fr} (NenuFAR, \citet{Zarka2012}) and the low frequency component of the Square Kilometre Array\footnote{https://www.skatelescope.org} (SKA-Low, see for instance \citet{Mellema2013}) will allow us to map the sky with HI regions through the redshifted 21-cm signal emitted during Reionization \citep{Mellema2015}. SKA will produce 2D tomographic images of the 21-cm emission of the EoR at many redshifts. Geometrical studies can therefore focus directly on the 21-cm fields, using for example the 21-cm power spectrum (see \citet{Kim2013,Choudhury2018,Seiler2019,Pagano2020}) or the 21-cm bispectrum (see \citet{Hutter2020}) to extract topological information, such as the size distribution of ionized bubbles.

In this study, we also propose to explore the EoR through its topology, but using the Discrete Persistent Structure Extractor, also called $\disperse$\footnote{http://www2.iap.fr/users/sousbie/web/html/indexd41d.html} \citep{Sousbie2011}. This code relies on the discrete Morse theory and persistent homology. 
By computing gradients and critical points of fields, it extracts geometrical properties, among which peaks, walls, filaments and voids for a 3D field.
The persistent homology is used in this algorithm to only focus on significant features and for instance takes into account the noise in data sets by providing a way to suppress it by tuning a persistence level.  $\disperse$ has been written in order to analyse astrophysical structures and especially the large scale density structures of the Universe. Since then, many studies have extracted and analysed the matter filaments with $\disperse$ from cosmological simulations (e.g. \citet{Singh2020,Galarraga-Espinosa2020,Katz2020,Song2021}), observations (e.g. \citet{Malavasi2020,Tanimura2020}) or both (e.g. \citet{Sousbie2011b}). Persistent homology has also been used on phenomenological models of the EoR to study the topology of HII regions \citep{Elbers2019}.

In this work, we extracted the filaments of the matter density field but also the peak patches of the redshift of reionization field or $z_{reion}$ \citep{Battaglia2013,Deparis2019} from EoR simulations produced by the semi-analytical code $\cmfast$\footnote{https://github.com/andreimesinger/21cmFAST} \citep{Muray2020,Mesinger2011} and the simulation code $\emma$ (Electromagnétisme et Mécanique sur Maille Adaptative, \citet{Aubert2015}). The $z_{reion}$ field is a measure of the redshift at which each point in space has been reionized. It contains therefore spatial and temporal information on the propagation of radiation from sources of reionization in a single field per simulation. 
Using $\disperse$, we apply the Morse theory framework to the 3D field of our different models. As detailed later on, maxima (peaks) in $z_{reion}$ correspond to the seeds, the first sources, of local reionizations. Peak patches around these seeds represent the extent of their time-integrated radiative influence. Even though we did not investigate them in the current work, walls or filaments in $z_{reion}$ define the regions where percolation occurred. We believe that such an approach can complement the ones based on the study of HII regions at a single or a sequence of redshifts. Possibly, the shape of these $z_{reion}$ patches can give us an indication on the way radiation get out of galaxies. Their size could inform us on the extent of the radiative influence of the different kinds of sources, in relation for example to the large absorption trough in the quasar spectra \citep{Becker2015} or to the understanding of the observed large fluctuations of UV radiation \citep{Chardin2017}. In general, one could expect a different long-range influence by faint/locally sub-dominant sources compared to more dominant emitters. Overall, we believe that the Morse theory framework as implemented by $\disperse$ provides a solid and reproducible mathematical description of the $z_{reion}$ geometry, that could be used to characterise and compare models of the EoR. In this initial study, we gain first insights on the extent of the $z_{reion}$ patches, their shapes and their relative orientation to the underlying gas distribution. We compare different EoR models from $\cmfast$ and $\emma$.

In this paper, we first describe in Sect. \ref{sec:methodology} the methodology we employ by presenting our simulations, the reionization redshift field, the $\disperse$ algorithm and the method we use to extract the orientation and shape of the patches. Afterwards, Sect. \ref{sec:results1} shows our first topological analyses: number of seeds of reionization, shape and size of the reionization patches. Sections \ref{sec:orientation} and \ref{sec:persistence} highlight the differences between simulations with different parameters, as well as studies of the orientation of radiation with respects to the underlying density. Section \ref{sec:EMMAsimus} presents a comparison of the topology of the EoR in semi-analytical and cosmological simulations. 
We finally conclude in Sect. \ref{sec:conclusion}.
In the whole paper, the cosmology parameters used are $(\Omega_m,\Omega_b,\Omega_{\Lambda},h,\sigma_8,n_s)=(0.31,0.05,0.69,0.68,0.81,0.97)$ as given by \citet{PlanckCollaboration2020}.

%--------------------------------------------------------------------
%--------------------------------------------------------------------
%--------------------------------------------------------------------
\section{Methodology}
\label{sec:methodology}
\subsection{Simulated data}

At first, we get 3D cosmological simulations from the semi-analytical code $\cmfast$ (version 3.0.3; 
\citet{Muray2020,Mesinger2011}). $\cmfast$ evolves initial density and velocity fields thanks to the first- or second- order perturbation theory of \citet{Zeldovich1970}, and provides associated predictions such as temperature, ionization, 21-cm signal and radiation fields. 

In our case, the ionized gas evolves in a $189^3$ cMpc\textsuperscript{3} box at a resolution of $1.48^3$ cMpc\textsuperscript{3} (or boxes of $128^3$ cMpc\textsuperscript{3}/h\textsuperscript{3} with $128^3$ cells). Different physical models are studied in this work using two parameters of \cmfast: the virial temperature $T_{vir}$ and the ionizing efficiency $\zeta$, as defined by \citet{Greig2015}:
\begin{itemize}[topsep=0pt,itemsep=0pt,parsep=0pt,partopsep=0pt]
    \item $T_{vir}$ is the minimal virial temperature for a halo to start the star-formation process. It allows to control the gas accretion, the cooling and the retainment of supernovae outflows, and it is related to the halo mass through $M_{min}\propto T_{vir}^{3/2}$. The higher the virial temperature, the heavier the halos.
    \item $\zeta$ is the ionizing efficiency of high-z galaxies. It defines the number of photons escaping from galaxies, and higher values tend to accelerate the Reionization.
\end{itemize}

\begin{figure}
   \centering
   \includegraphics[width=0.5\textwidth]{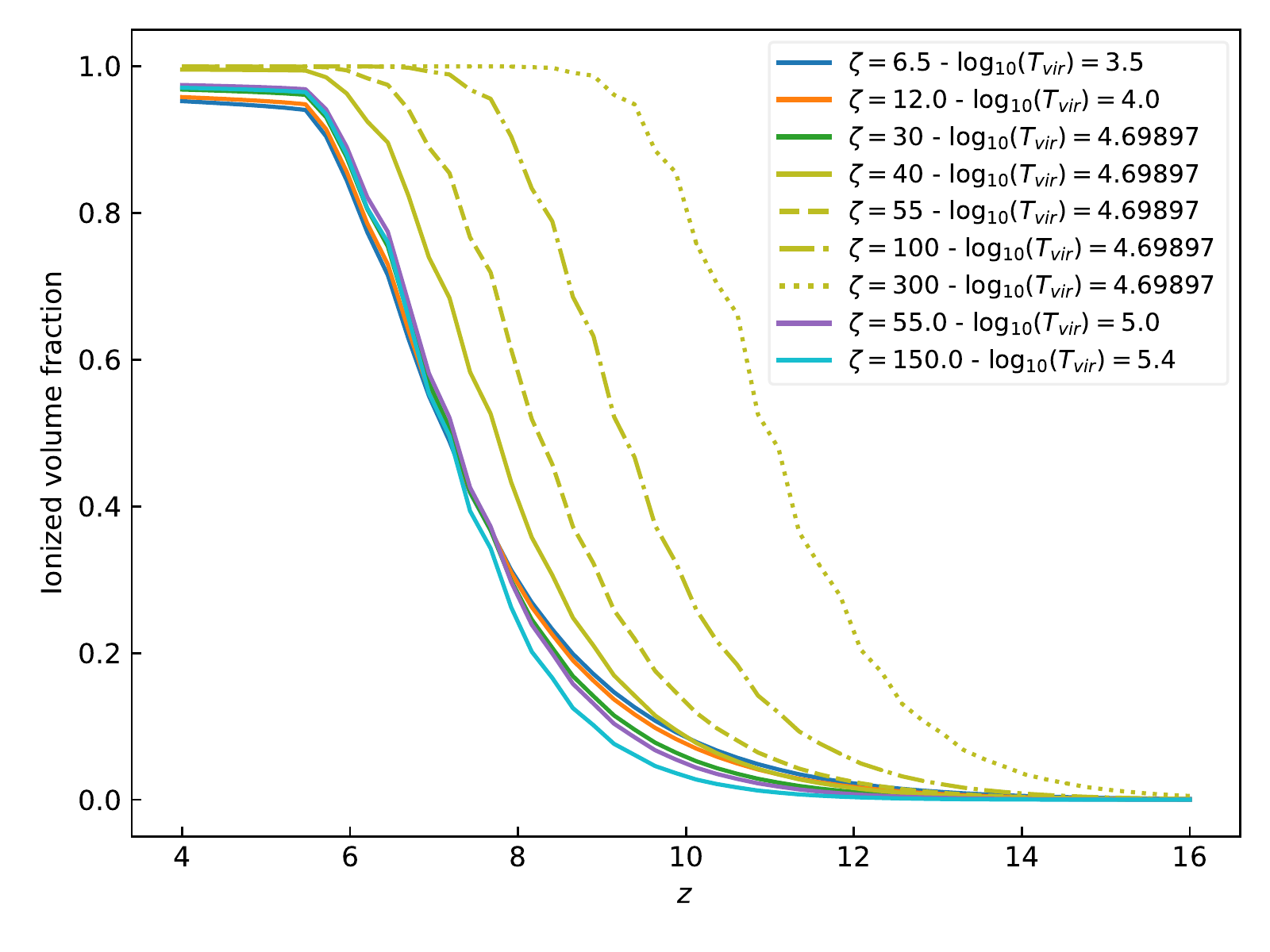}
    \caption{Ionized volume fraction of sets of $\cmfast$ simulations having different models, each one of them averaged on their 101 realisations.}
    \label{fig:xion_21cmfast}
\end{figure}

The parameters of our different models are shown in Table \ref{table:21cmFASTparams}. We separate the models into two sets of simulations: a set varying $\zeta$ and $T_{vir}$, called the Same Reionization History (SRH) set, and a set varying only $\zeta$, called the Different Reonization History (DRH) set. For the SRH simulations set, virial temperatures have been set first before tuning the ionizing efficiency in order to get the same history of reionization. Then, the larger $T_{vir}$ is, the larger $\zeta$ is in order to maintain the same history of reionization since only the heaviest halos radiate, meaning that they are also fewer to release photons. The DRH simulations set have been made in order to analyse the impact of different emissivities and thus reionization histories. 101 realisations of each model have been made using different seeds in $\cmfast$. The ionization histories of Fig. \ref{fig:xion_21cmfast} are averaged on all the realisations of each model. Figure \ref{fig:xion_21cmfast} shows the ionized volume fraction of all models. 

\begin{table}   
\centering                          
\begin{tabular}{c c c c c}       
\hline\hline                 
Model & $\zeta$ & $T_{vir}$ [K] & \footnotesize{\textit{($\log_{10}(T_{vir})$)}} \\    
\hline                        
    1   & 6.5 & $3.2\times 10^{3}$ & \footnotesize{\textit{(3.5)}} \\      
    2   & 12  & $1\times 10^{4}$ & \footnotesize{\textit{(4)}} \\
    3   & 30  & $5\times 10^{4}$ & \footnotesize{\textit{(4.69897)}} \\
    4   & 55  & $1\times 10^{5}$ & \footnotesize{\textit{(5)}} \\
    5   & 150 & $2.5\times 10^{5}$ & \footnotesize{\textit{(5.4)}} \\ 
\hline
    3   & 30  & $5\times 10^{4}$ & \footnotesize{\textit{(4.69897)}} \\
    3-1 & 40  & $5\times 10^{4}$ & \footnotesize{\textit{(4.69897)}} \\
    3-2 & 55  & $5\times 10^{4}$ & \footnotesize{\textit{(4.69897)}} \\ 
    3-3 & 100  & $5\times 10^{4}$ & \footnotesize{\textit{(4.69897)}} \\ 
    3-4 & 300  & $5\times 10^{4}$ & \footnotesize{\textit{(4.69897)}} \\ 
\hline                                   
\end{tabular}
\caption{Parameters of the $\cmfast$ simulated models. There are five models that change $\zeta$ and $T_{vir}$ that belong with the Same Reionization History (SRH) set, and four other models modifying the $\zeta$ parameter from the fiducial model \#3, which form the Different Reionization History (DRH) set. The values of $T_{vir}$ are also given in logarithm.}       
\label{table:21cmFASTparams}      
\end{table}

\subsection{Reionization redshift maps}

As mentioned in the Introduction, the geometry of ionization is usually investigated via the ionized fraction distribution $x_{HII}(\vec r,z)$ at a given redshift $z$ or with a sequence of redshifts, where $\vec r$ stands for the 3D position. Here, we rather focus on the reionization redshift field $z_{reion}$. It is obtained by saving for each position the redshift at which the gas is considered to be ionized, that is when the ionization fraction gets higher than a given threshold which is set at 50\%\footnote{A few cells are not reionized at the end of the simulations ($z=5.5$). The reionization redshift of those cells is manually set to 5.5 in the $z_{reion}$ map.}, as follows: 
\begin{equation}
    z_{reion}(\vec r)=z(\vec r,x_{HII}=0.5).
\end{equation}
It incorporates spatial and temporal information on the propagation of radiation during the Epoch of Reionization in a single field \citep{Battaglia2013,Deparis2019,Trac2021} and can be seen as a reciprocal of or dual to the $x_{HII}$ field.

\begin{figure}
   \centering
   \includegraphics[width=0.5\textwidth]{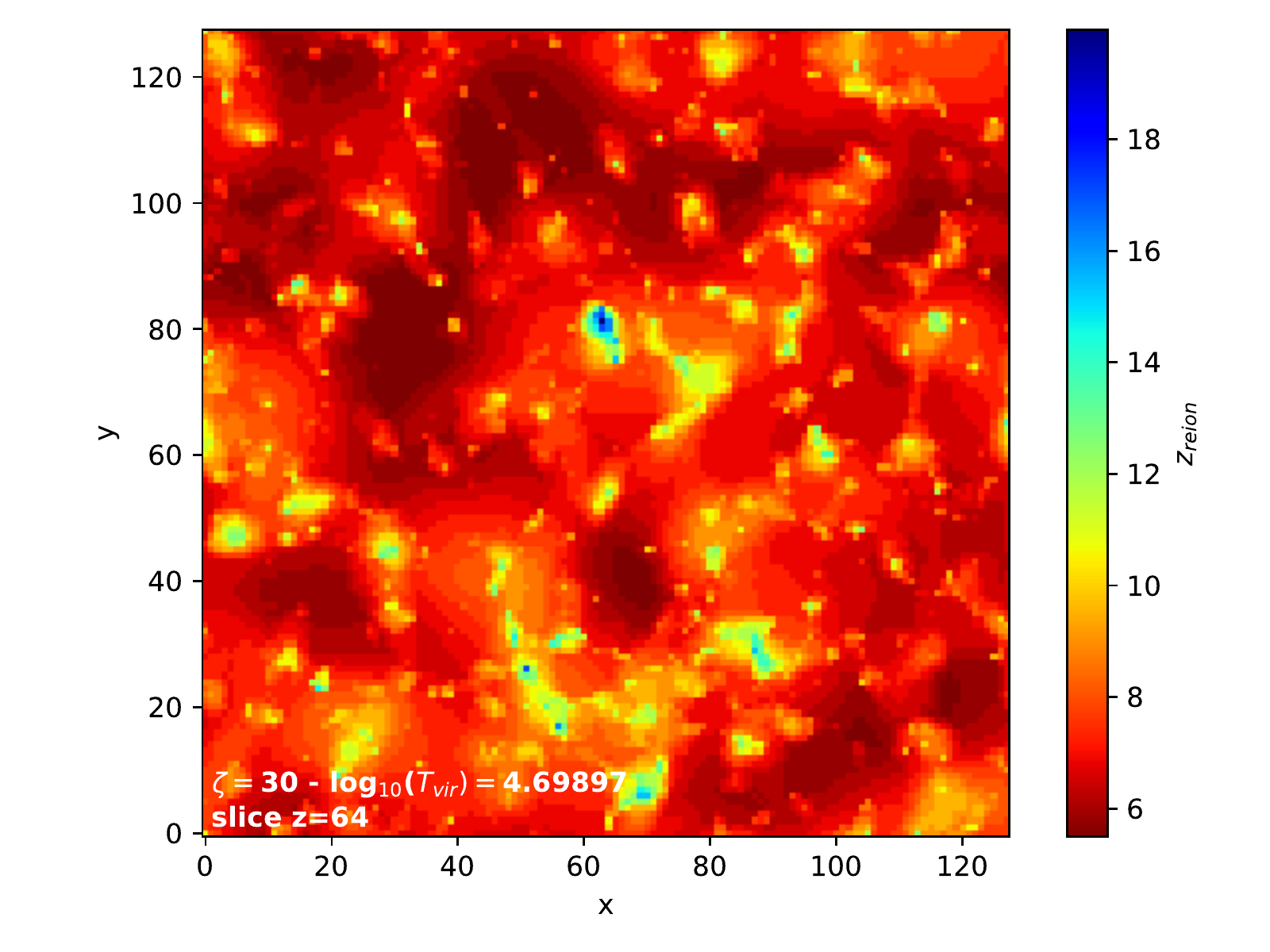}
    \caption{2D slice at depth $z=64$ of the reionization redshift field extracted from the $\cmfast$ simulation (model $\zeta=30$ and $T_{vir}=5\times 10^{4}$ K).}
    \label{fig:21cmFAST12345_exZreion}
\end{figure}

Figure \ref{fig:21cmFAST12345_exZreion} shows an example of a reionization redshift map extracted from a $\cmfast$ simulation. The bluest regions correspond to the places where the gas was reionized the first and can be interpreted as the local seeds of the local reionization process. Ionization fronts propagated around those $z_{reion}$ peaks towards the reddest regions which have reionized later on, forming connected regions of varying shapes and extent: these regions represent the local radiative influence of these local seeds, integrated over time.

\subsection{$\disperse$}
\label{sec:disperse}

$\disperse$ relies on the Morse theory in order to analyse the topology of manifolds through the study of differential functions. This code searches for critical points in a field, such as maxima, minima and saddle points. 
$\disperse$ also looks for the integral lines (or field lines) of the field: they are the tangent curves to the gradient in every point, and always have critical points as origin and destination. As those lines cover all space, one can produce a tessellation of the space, creating regions called ascending or descending manifolds. 
Finally, the set of all ascending/descending manifolds (also known as peak or void patches) is called the Morse complex of a field. 

\begin{figure}
   \centering
   \includegraphics[width=0.5\textwidth]{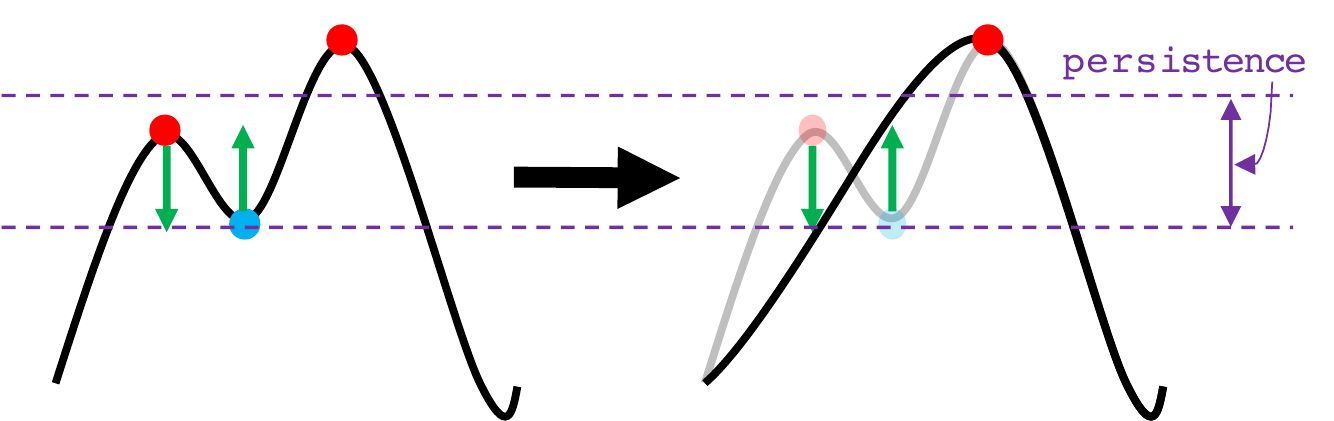}
    \caption{Illustration of the persistence parameter of the $\disperse$ algorithm. It shows a 1D function, with red dots as maxima and blue dots as minima. The left column is the actual function on which the persistence threshold is applied, while on the right one, there is the result after applying this threshold. The length between the purple dashed lines represents the persistence threshold in that case. The green arrows represent the smoothing of the curve when eliminating the corresponding critical points. This illustration is inspired by \citet{Sousbie2011}.}
    \label{fig:persistence_illustration}
\end{figure}

Within $\disperse$, one can tune the so-called "persistence" to control the significance of the topological features found in fields (see \citet{Sousbie2011}). It can be seen as a significance threshold that separates two critical points, and the local critical points of the field which are not significant enough are ignored. It allows to control the smoothness of the resulting topological features and can be used to get rid of the noise in the input data.
Figure \ref{fig:persistence_illustration} illustrates the impact of applying a persistence threshold on a 1D field (left column) on which a persistence threshold is applied (right column). Let’s consider a pair of critical points: a minimum (in blue) and a maximum (in red). The distance between the value of the field at those points (green arrows) is compared to the value of the chosen persistence threshold (distance between the two purple dashed lines): if it is larger than the persistence, those extrema are considered when detecting and assigning points to patches; but if it is smaller, they are discarded, as if they were filtered out from the field. In this case the field becomes topologically "smoother". 

Before using $\disperse$, the gas density $\delta(\vec r)$  and the reionization redshift $z_{reion}(\vec r)$ are first converted into dimensionless 3D fields by applying the following transformation on them:
\begin{equation}
    \delta_{a} = \frac{\delta-\overline{\delta}}{\sigma_{\delta}}, 
\end{equation}
\begin{equation}
    z_{reion,a} = \frac{z_{reion}-\overline{z_{reion}}}{\sigma_{z_{reion}}}, 
\end{equation}
where $\overline{\delta}$ and $\overline{z_{reion}}$ are the mean of each field and $\sigma_{\delta}$ and $\sigma_{z_{reion}}$ are their standard deviation. These transformations allow us to express the persistence parameter in term of the deviation of a field to its mean value, letting us know at how many $\sigma$ the fields will be smoothed from a topological point of view. Moreover, the fields are also filtered with a Gaussian kernel (with a standard deviation of 1 cell for each field) in order to avoid the presence of large patches of constant field values that prevent $\disperse$ to run properly.

\begin{figure*}
   \centering
   \includegraphics[width=1\textwidth]{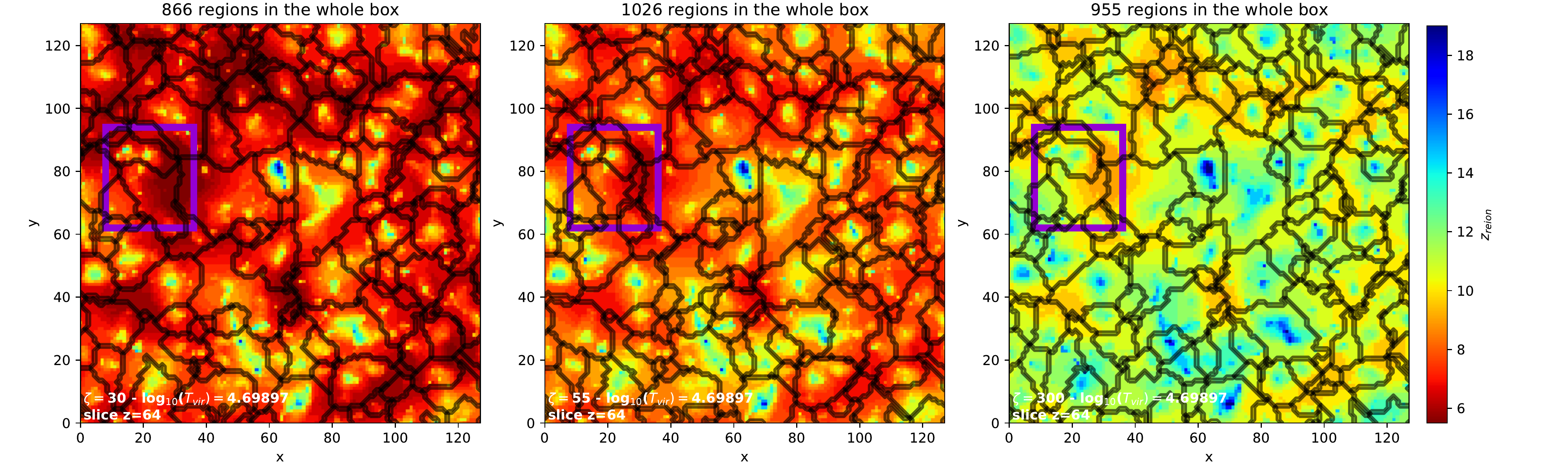}
    \caption{2D slices of $z_{reion}$ fields extracted from a realisation of $\cmfast$ simulations. The reionization redshifts are superimposed to the contours (in black) of the patches from the segmentation obtained with $\disperse$ (with a persistence level of $0.5-\sigma$). The three panels represent three models (of the DRH simulations set) with different $\zeta$ value: models \#3 ($\zeta=30$), \#3-2 ($\zeta=55$) and \#3-4 ($\zeta=300$) from left to right.}
    \label{fig:ex_zreionRegs}
\end{figure*}

In our study, we first apply the $\disperse$ algorithm with a persistence level of $0.5-\sigma$ \citep{Sousbie2011b,Codis2018,Galarraga-Espinosa2020,Cohn2021} on the 3D reionization redshift field of our simulations in order to get the $z_{reion}$ patches associated with the $z_{reion}$ maxima\footnote{In practice, $\disperse$ compute the ascending 3-manifolds of the field using the integral lines that are related to the maxima of the 3D field (see \citet{Sousbie2011} for more details).}. They correspond to the regions that are topologically associated to a given $z_{reion}$ maximum, which gather all the cells of the simulations that have a positive gradient toward their maximum. Those maxima are the first places to reionize: they are the seeds of local reionizations.  
Again, these patches are different from the "ionized bubbles" usually mentioned in the literature (see e.g. \citet{Chen2019,Gorce2019,Giri2019a,Giri2020}), that are regions where the hydrogen gas is ionized at a given time. The cells within the patches discussed here are connected \textit{through} time via the topology of $z_{reion}$, providing another insight on the propagation of ionization fronts and on the extent of the influence of the reionization seeds.
Figure \ref{fig:ex_zreionRegs} shows reionization redshift maps superimposed to the contours of their segmentation obtained with $\disperse$. One can see that the patches are naturally centred around the maxima, not necessarily spherical, and have different kinds of sizes and shapes (keeping in mind that 3D effects can be hidden on these 2D maps) depending on the local specificities of the production and absorption of radiation. 

\begin{figure}
   \centering
   \includegraphics[width=0.5\textwidth]{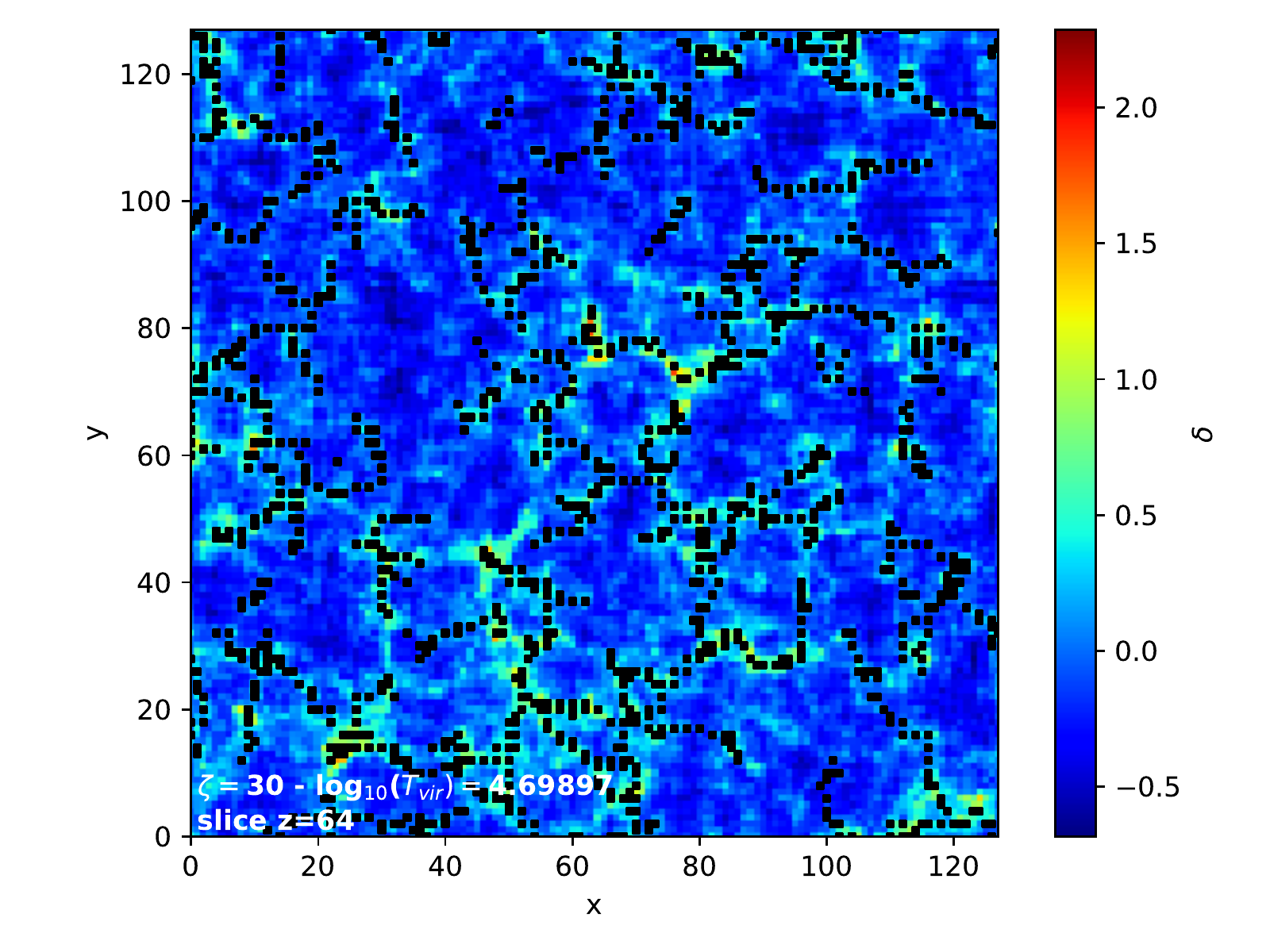}
    \caption{2D slice of the density field extracted from a $\cmfast$ simulation (model $\zeta=30$ and $T_{vir}=5\times 10^{4}$ K). The density is superimposed to the filaments (in black) obtained with $\disperse$ (with a persistence level of $0.5-\sigma$). This map is made from an average of many slices to better visualise the filaments.}
    \label{fig:ex_densFils}
\end{figure}

We are also interested in the filaments of the gas density structures present in our models. With $\disperse$, they are detected as the set of points/cells in the density field belonging to arcs that have saddle points as origin and extrema as destination. We apply this selection to the gas density fields in our simulations using a $0.5-\sigma$ persistence level. Figure \ref{fig:ex_densFils} is an example of a density map from a $\cmfast$ simulation superimposed to the filaments detected with $\disperse$. The map is obtained from an average of several slices for a better visualisation and clearly the filaments follow the distribution of the gas in the box.

\subsection{The inertial tensor, a tool to get the orientation of an object}
\label{sec:inertialTensor}

Our study will assess the shape and size of the reionization patches, as well as their orientation with respect to the gas filaments. We use inertial tensors, as defined by \citet{Tormen1997}: 
\begin{equation}
    I_{i,j} = \sum_k \omega_k \cdot (x_{k,i}-x_{ref,i}) \cdot (x_{k,j}-x_{ref,j}).
\end{equation}
Here, $k$ scans all of the cells within the patches, $i,j \in \{1,2,3\}$ refer to the Cartesian components of the cells position $x$ and $\omega$ is a weight. $x_{ref}$ is a point of reference: in our case, it is the geometrical centre of each patch. The eigenvectors of the tensor provide the 3D directions of the extension of the patches and the eingenvalues the typical extent of the cells along these directions. The eigenvector with the largest eigenvalue represents therefore the main axis of a given patch. The same procedure can be applied to the cells belonging to a given gas filament, in order to extract its main orientation. 

For $z_{reion}$ patches, we use $\omega_k=1$, as we are only interested in the geometrical shape, and their main axis represents the direction favoured by radiation after having been emitted by the local seed. Presumably, it corresponds to the path of least resistance for radiation or is indicative of the geometry of light production as ionization fronts propagated. For gas filaments, since we are interested in their orientation relatively to the reionization patch they are located in, we set $\omega_k$ to the cell density and restrict ourselves to cells that are within that reionization patch.

%--------------------------------------------------------------------
%--------------------------------------------------------------------
%--------------------------------------------------------------------
\section{First topological insights}
\label{sec:results1}

\subsection{Number of reionization patches}
\label{sec:numbSources}

\begin{figure}

   \centering
   \includegraphics[width=0.5\textwidth]{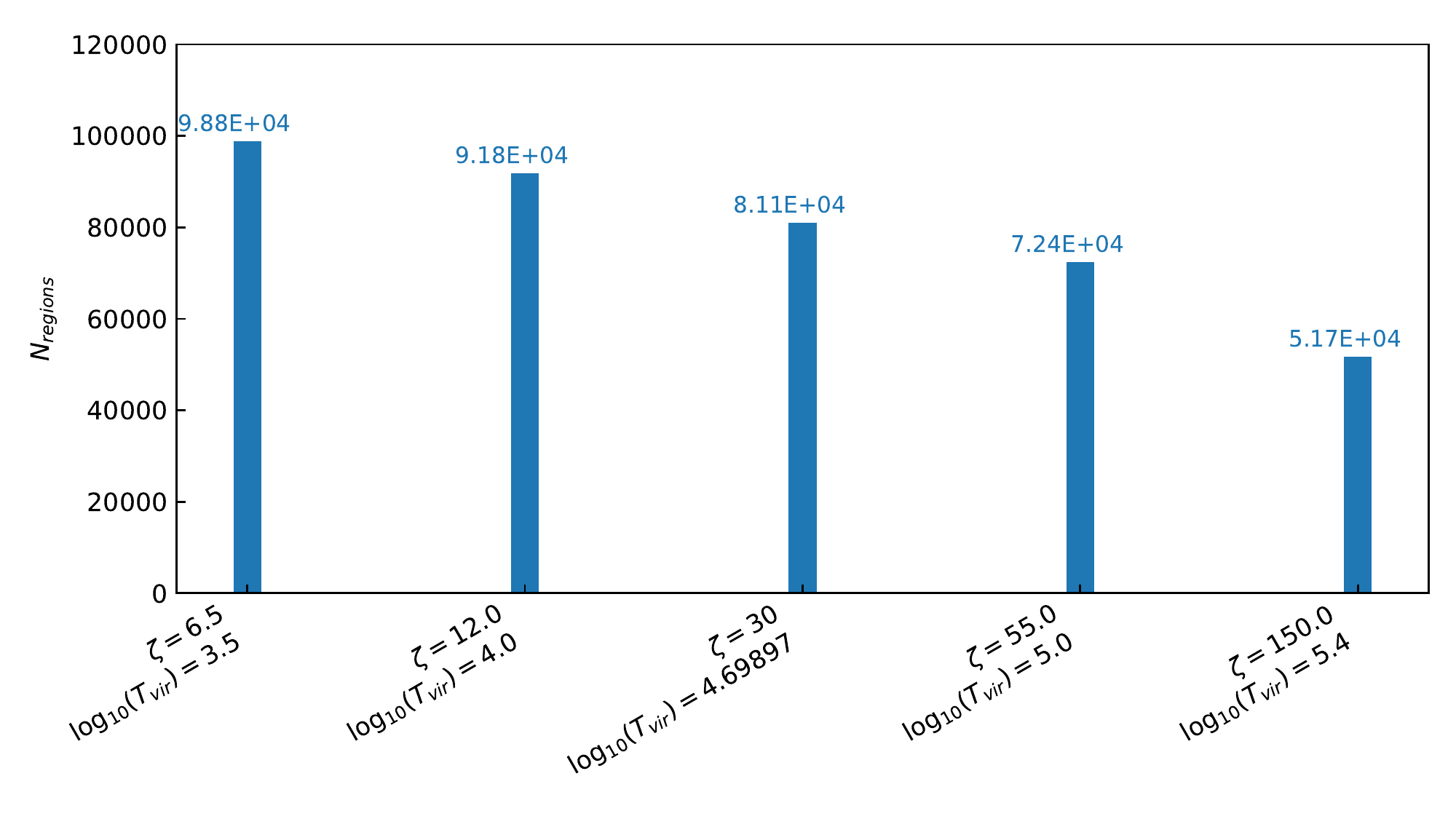}
   \includegraphics[width=0.5\textwidth]{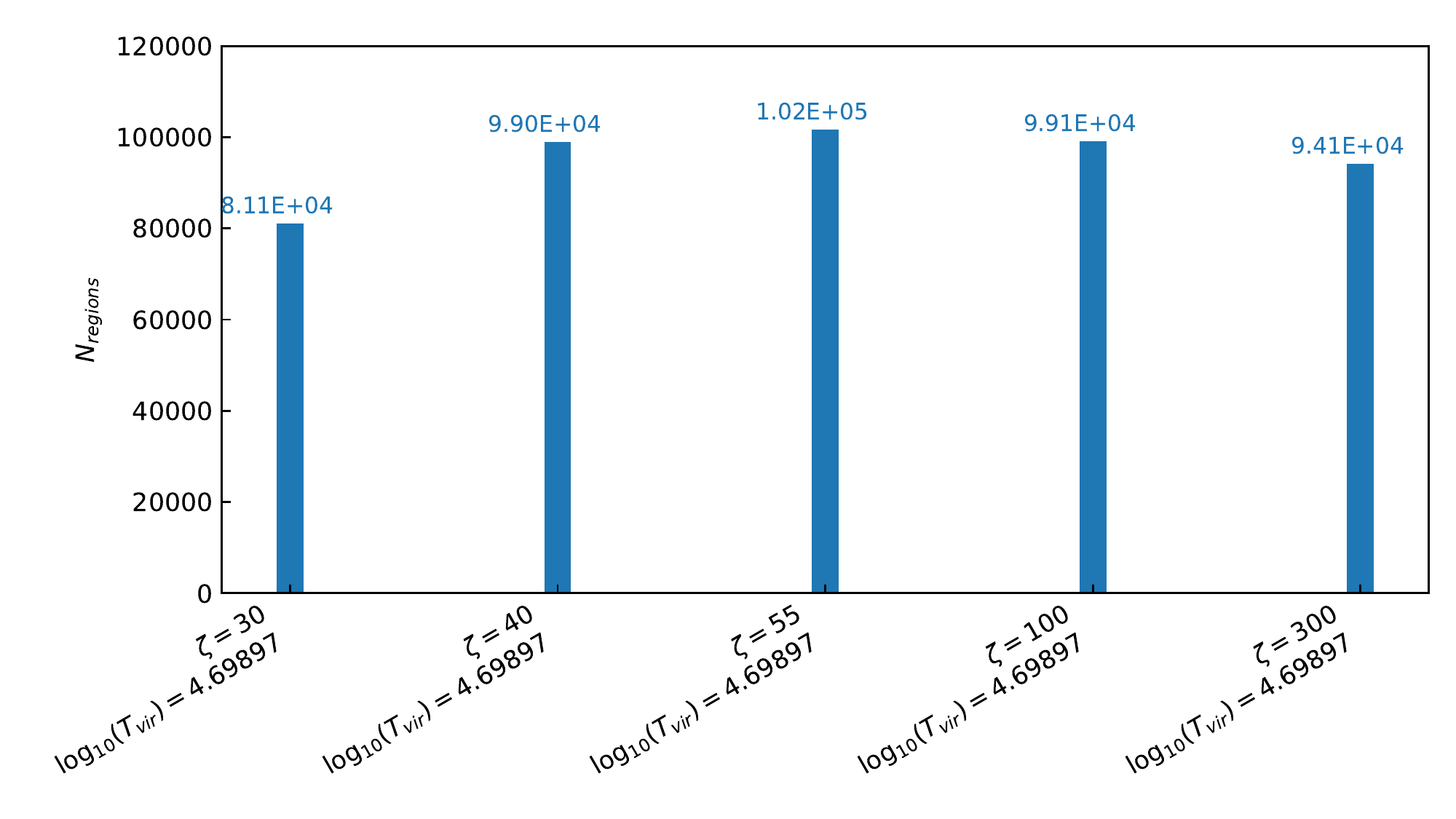}
    \caption{Number of patches detected with $\disperse$ (with a persistence level of $0.5-\sigma$) for every model mentioned in Table \ref{table:21cmFASTparams}. The top panel represents the models of the SRH set while the bottom panel shows the models of the DRH set. For each model, one adds the patches numbers of every realisation.} 
    \label{fig:21cmfast_NregFil}
\end{figure}

On Fig. \ref{fig:21cmfast_NregFil}, one can see the number of reionization patches in every model (accumulating all realisations). It should be noted that the parameters evolving from one simulation to the other, $T_{vir}$ and $\zeta$, only have an impact on the radiation processes whereas the density maps are not modified between the different $(T_{vir},\zeta)$ models, and for instance the filaments are necessarily the same. One can see that the number of reionization patches (or reionization seeds) decreases with increasing $T_{vir}$ (and then $\zeta$, see the top panel of Fig. \ref{fig:21cmfast_NregFil} representing the SRH set): with larger $T_{vir}$, there are less emitting halos as it favours the massive ones (which also produce more photons to maintain a reionization with the same history). 
Now, for the DRH simulations set (i.e. with the same emitting halos but different ionizing efficiencies), the number of reionization patches as a function of the ionizing efficiency parameter presents two regimes. We would have expected the number of reionization patches to only decrease with increasing $\zeta$: when the sources emit more radiation, the first regions expand rapidly, preventing the smaller and later-formed halos to create their own patches. The latter sources are then ionized by external emitters, and belong to the patches of these earlier external "reionizers".
For instance, this behaviour can be seen in the purple boxed patch in the middle and right panels of Fig. \ref{fig:ex_zreionRegs}.  It shows reionization redshift fields for $\zeta$ values of 55 and 300 superimposed to their $\disperse$ segmentations. As $\zeta$ increases, four small patches merge as bright early emitters expand their influence more efficiently and late emitters are externally reionized.
However, the number of patches increases as $\zeta$ increases from 30 to 55. We checked that this behaviour is an artefact of the way the $\disperse$ segmentations are produced in this study. The $z_{reion}$ field is smoothed before being processed by $\disperse$, filtering out small patches of low emitters with weak gradients. As $\zeta$ increases, the extent and the gradients are enhanced within such patches and the latter are not suppressed by the Gaussian smoothing and the persistence filter.
This effect is illustrated in the left and middle panels of Fig. \ref{fig:ex_zreionRegs}. Only two patches are detected within the purple box at low $\zeta$, while four patches can be seen as this parameter is increased to moderately larger values. 

The number of patches is related to the first Betti number $\beta_0$ of an ionization field as it represents the number of sources of ionization \citep{Giri2020}. \citet{Giri2020} compute the $\beta_0$ number of their $x_{HII}$ ionization field throughout the Reionization process for models that vary only the ionizing efficiency (see their FN1 and FN2 models and the top left panel of their Fig. 6 where they accumulate the number of sources as $x_{HII}$ grows). Those models can be compared to our DRH set models for which we have histories of reionization (for the lowest $\zeta$ models) similar to theirs. They conclude that the $\beta_0$ numbers of these two models are approximately the same, but with slightly less sources, for the model with the highest $\zeta$. Meanwhile, we conclude that reionization patches appear sooner (see Fig. \ref{fig:xion_21cmfast}) and their total number throughout the reionization history is lower for the highest $\zeta$ models, (see Fig. \ref{fig:21cmfast_NregFil})  which is consistent with the results of \citet{Giri2020}.

\subsection{Shapes of the reionization patches}

\begin{figure}
   \centering
   \includegraphics[width=0.5\textwidth]{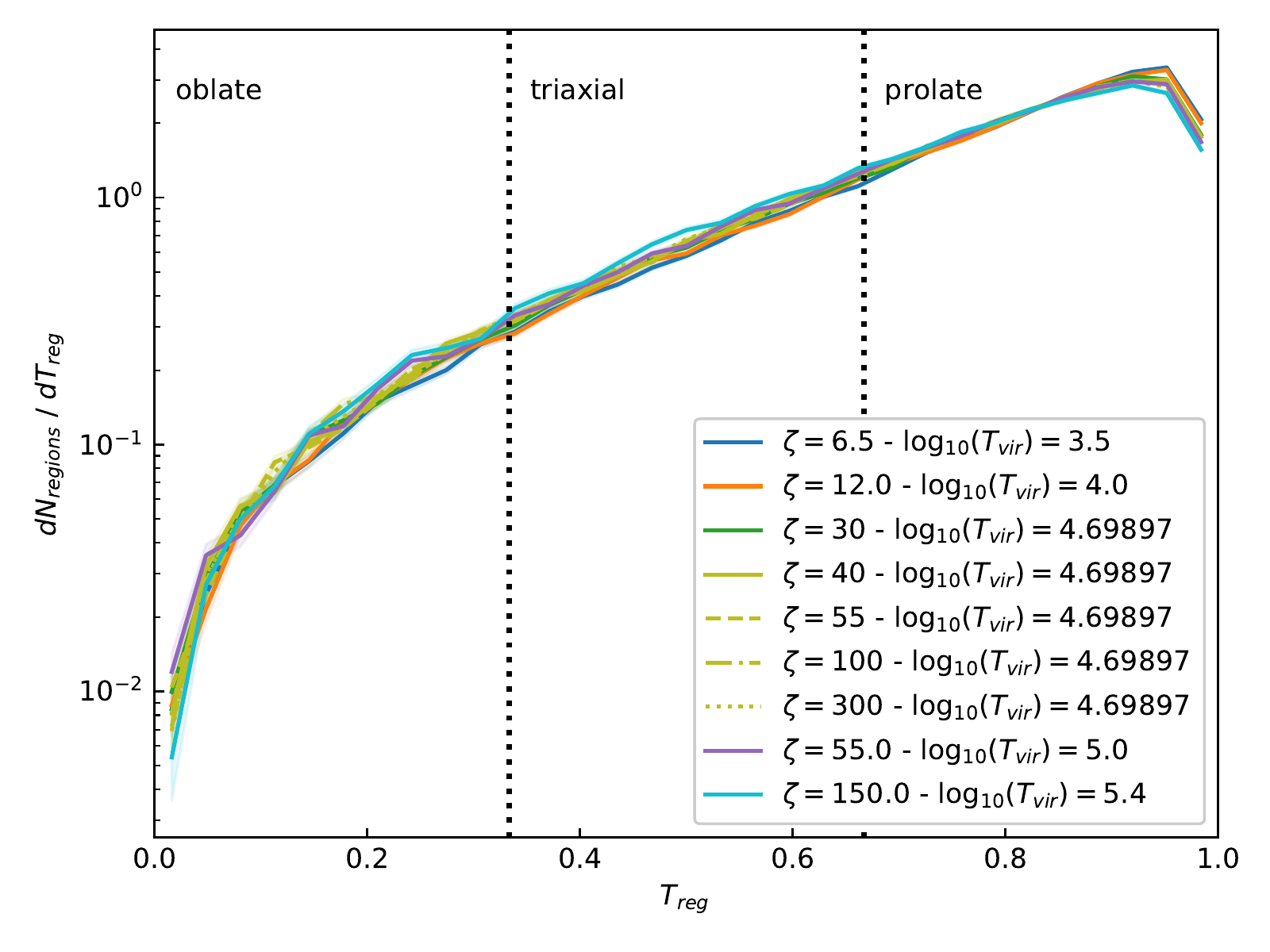}
   \includegraphics[width=0.5\textwidth]{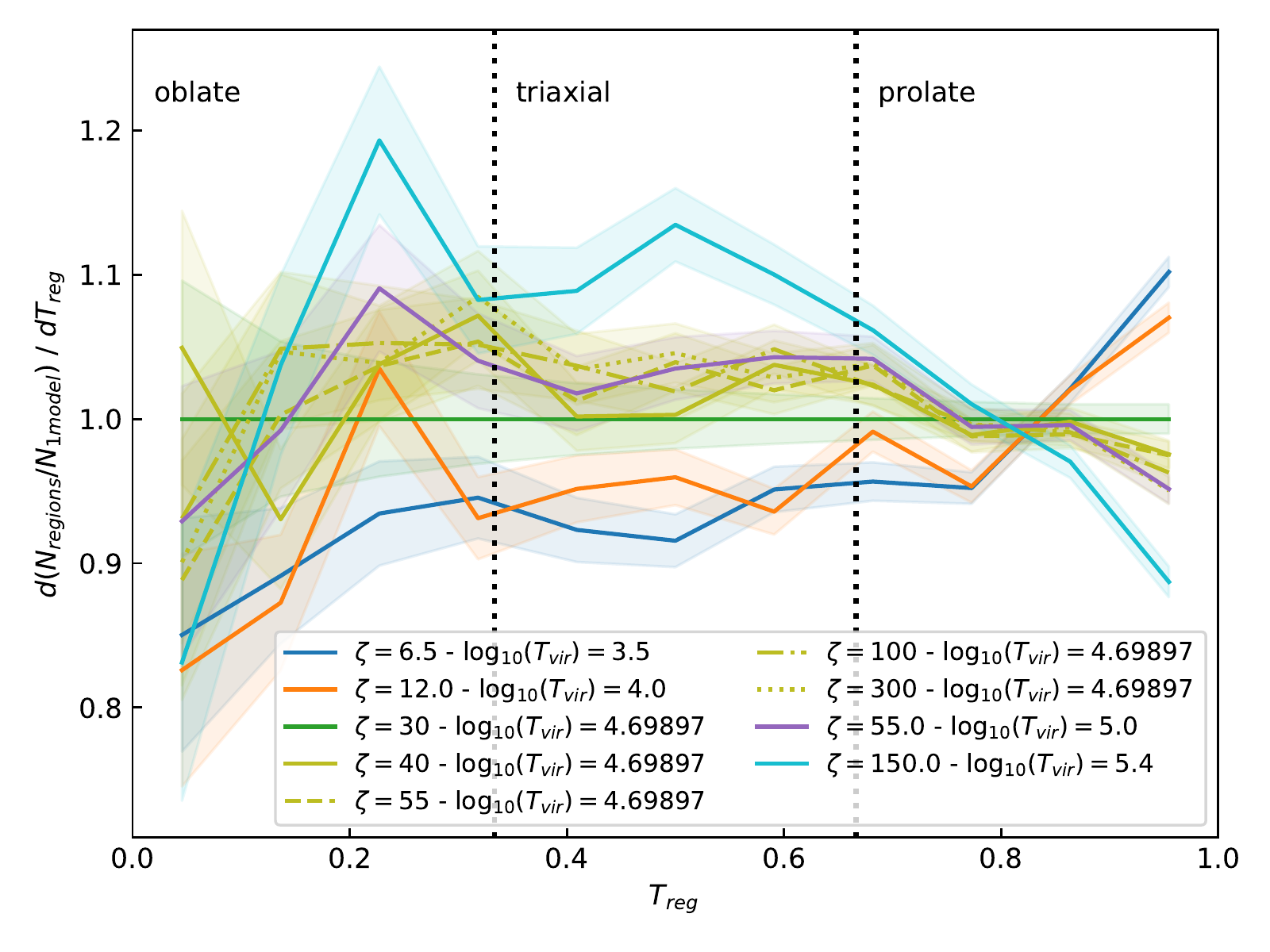}
    \caption{Probability distribution functions of the triaxiality parameter of the reionization patches of every model (top panel). The bottom pannel represents the same PDFs but each one of them is divided by one particular model (the one with $\zeta=30$ and $\log_{10}(T_{vir})=4.69897$).  On the two panels, all the patches of each realisation for one specific model are accumulated.}
    \label{fig:21cmfast_T}
\end{figure}

With the eigenvalues given by the inertial tensors, one can compute a number quantifying the geometrical shape of the patches. \citet{Tormen1997} called it the triaxiality parameter and defined it as follows:
\begin{equation}
    T = \frac{\lambda_3^2 - \lambda_2^2}{\lambda_3^2 - \lambda_1^2}.
\end{equation}
The $\lambda_i$ (with $i\in\{1,2,3\}$) are the eigenvalues of the inertial tensors with $\lambda_1\leq\lambda_2\leq\lambda_3$. When $0\leq T\leq\frac{1}{3}$, the object is oblate: it has two large dimensions and one small, like a flattened sphere. When $\frac{1}{3}\leq T\leq\frac{2}{3}$, the object is "triaxial". When $\frac{2}{3}\leq T\leq 1$, the object is prolate: it has one large dimension and two small ones, like a rugby ball. The top panel of Fig. \ref{fig:21cmfast_T} shows the probability distribution functions of the triaxiality parameters computed for the reionization patches of each realisation of every model. They all return a similar distribution with a majority of prolate patches and a minority of oblate ones. The normalisation of  these distributions with the fiducial model (bottom panel) reveals small differences between the SRH models though: the more $T_{vir}$ and $\zeta$ increase, the less there are prolate patches. 

These prolate or "cigar" shapes of the reionization patches is reminiscent of the "Ionized Fibres" stage as it is called in \citet{Chen2019}. Using Minkowski Functionals they could separate in five stages the process of Reionization: an "Ionized Bubble" stage where the first isolated ionized bubbles appear, an "Ionized Fibres" stage where bubbles start to connect to each other forming a large fibre structure, a "Sponge" stage with intertwined ionized and neutral regions, and "Neutral Fibre" and "Neutral Islands" stages, which are the neutral counterpart of the two first stages. In our segmentations, patches seem to have kept a memory of the fibre shape of the second stage of \citet{Chen2019}.

\subsection{Size distribution of the reionization patches}

\begin{figure}
   \centering
   \includegraphics[width=0.5\textwidth]{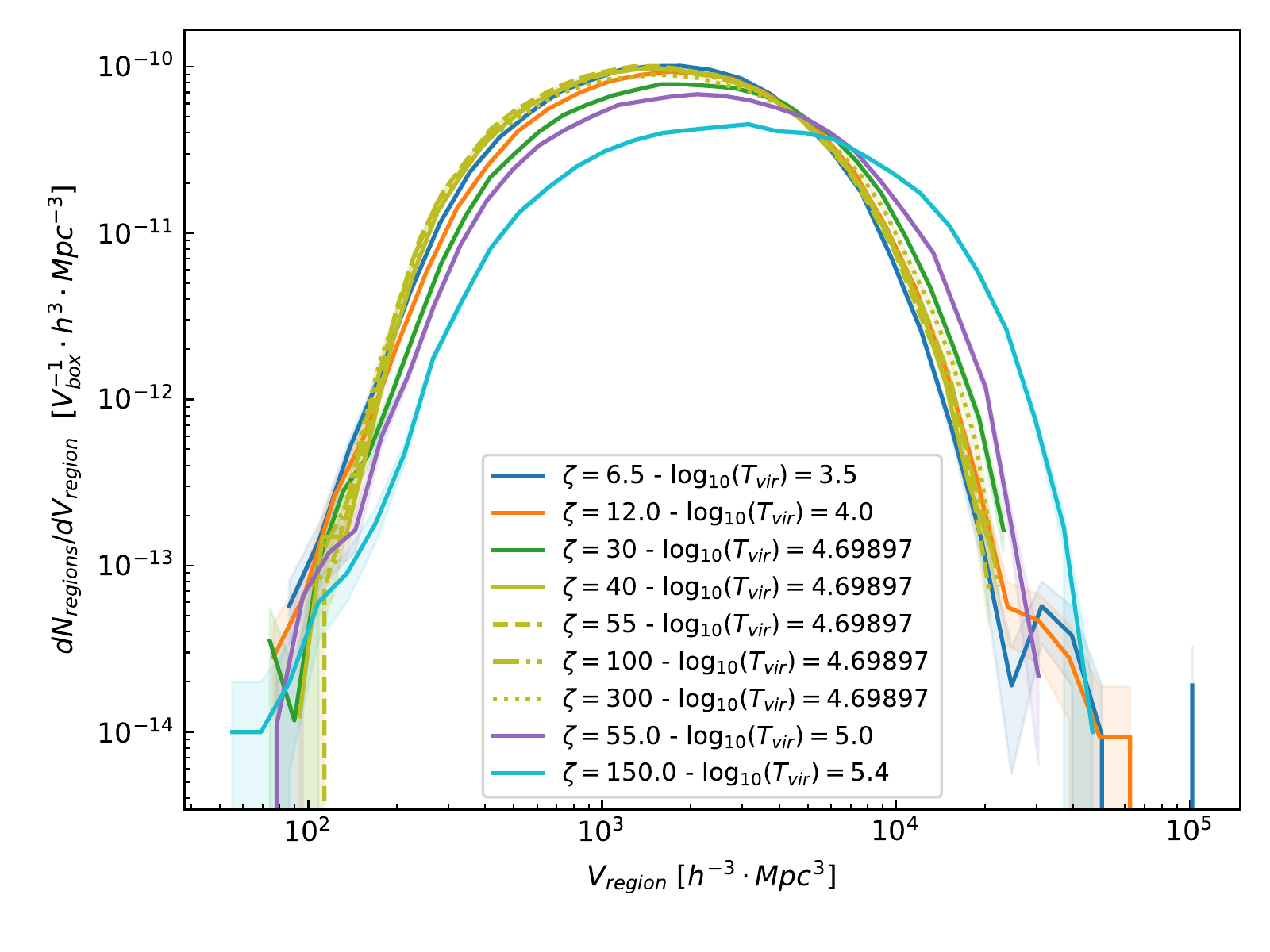}
    \caption{Probability distribution functions of the volume of segmentations patches for each physical model mentioned in Table \ref{table:21cmFASTparams}. Here, all the patches of each realisation for one specific model are accumulated.}
    \label{fig:distribSeg_21cmfast}
\end{figure}

We compute the probability distribution function of the volume of the reionization patches. Figure \ref{fig:distribSeg_21cmfast} shows those PDFs for the two sets of simulations. One can notice that patches with intermediate volumes dominate the simulation box. There is indeed a limit on the largest volume due to the finite size of the box and on the smallest volumes because of the resolution. Comparing the different models of the SRH set, the largest virial temperatures favour the ionization by the more massive halos that are stronger emitters: it leads to larger but fewer reionization patches. A similar trend is measured for the DRH set, when $\zeta$ becomes large, with fewer and more prominent patches.
From our size distributions, one can extract a typical "radius" of patches of $\sim10$ cMpc/h, consistent for instance with the radii found by \citet{Gorce2019} with their Triangle Correlation Function.

The size distribution of patches returns a general information of the maximal radius HII regions can reach before percolation during the whole reionization process. It thus provides insights about the required size for cosmological simulations of the EoR to use the framework described in this work. For example, our PDFs of the patches volumes indicate that our boxes ($128$ cMpc/h) do not limit typical-sized patches ($10$ cMpc/h), however it is not clear yet what kind of cut-off is applied to these PDFs in the regime of large reionization patches. It is for instance known that many features in the 21-cm signal only converges at larger scales (200 Mpc/h at least, \citet{Iliev2014}) and further investigations will be conducted later on to see how it translates to our reionization patches.

\section{Orientation of the reionization patches with respect to the gas filaments}
\label{sec:orientation}

This section deals with the orientation of the reionization patches with respect to matter. The directions of the main axis of the gas filaments and the patches they are located in are compared using the eigenvectors of their inertial tensors, computed as mentioned in Sect. \ref{sec:inertialTensor}. To do so, the cosine of the angle between those directions is computed as follows:
\begin{eqnarray}
    \cos(\langle z_{reion},\delta\rangle) & = & \cos(\langle \Vec{\Psi}_{z_{reion},3},\Vec{\Psi}_{\delta,3}\rangle), \\
    & = & \frac{\Vec{\Psi}_{z_{reion},3} \cdot \Vec{\Psi}_{\delta,3} }{||\Vec{\Psi}_{z_{reion},3}|| ||\Vec{\Psi}_{\delta,3}||},
\end{eqnarray}
$\Vec{\Psi}_{z_{reion},3}$ and $\Vec{\Psi}_{\delta,3}$ are the eigenvectors corresponding to the largest eigenvalue of the $z_{reion}$ patches and the filaments respectively.

\subsection{Emitting halos with different mass range and ionizing efficiency (SRH set)}
\label{sec:CosZetaTvir}

\begin{figure}
   \centering
   \includegraphics[width=0.5\textwidth]{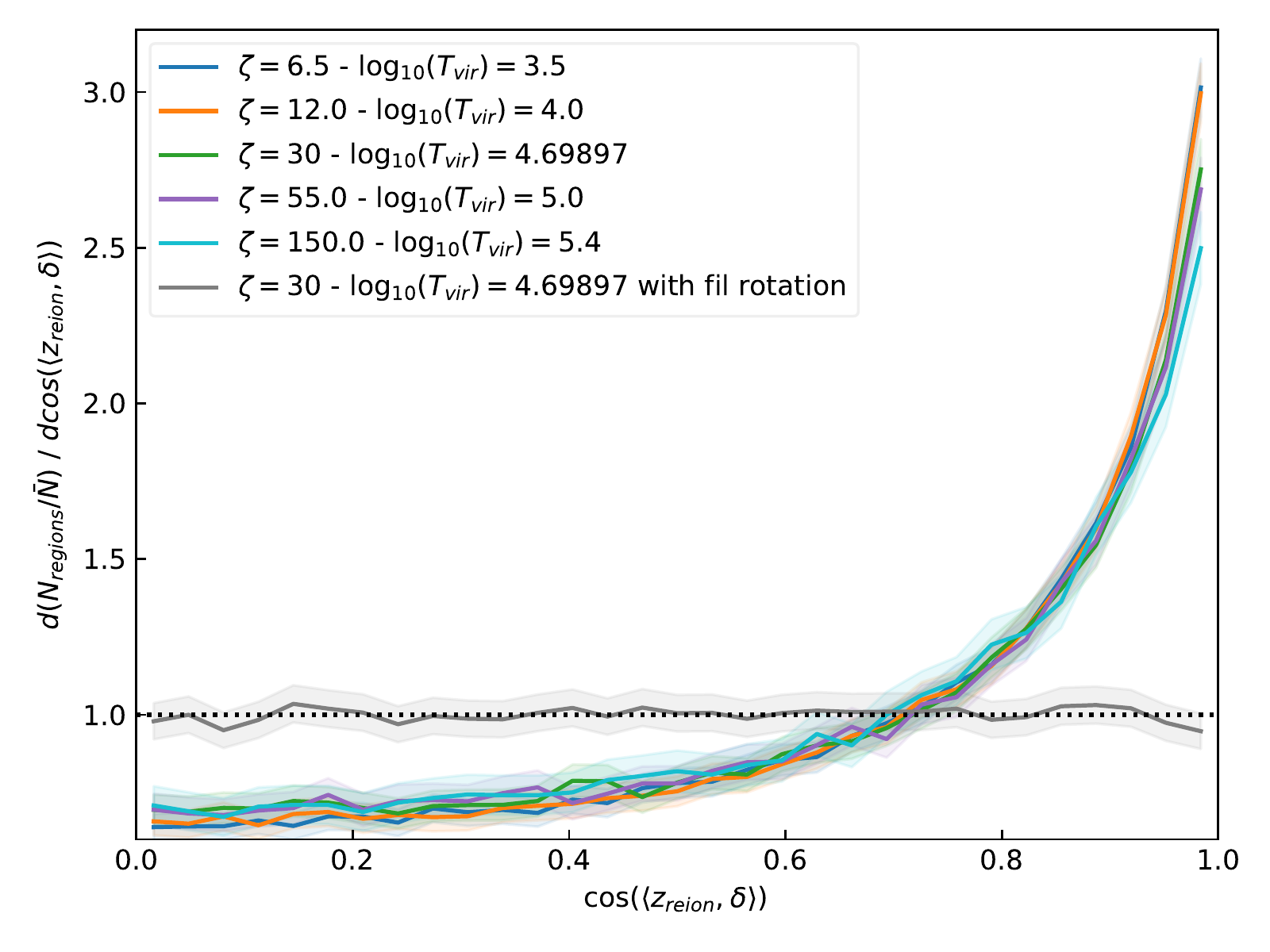}
   \includegraphics[width=0.5\textwidth]{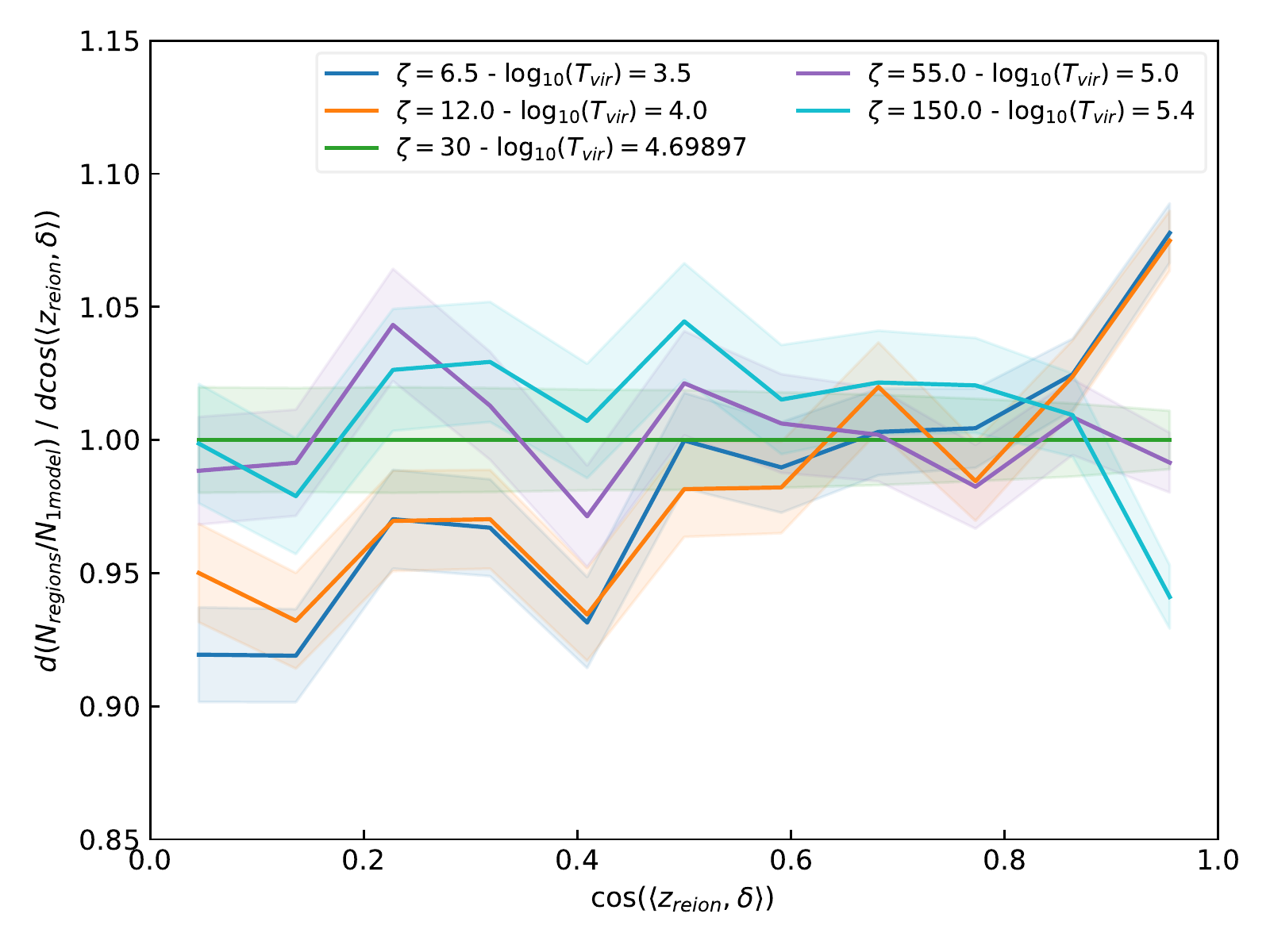}
    \caption{On the top panel, there are the probability distribution functions of the cosine of the angle between the reionization patches and their corresponding filaments (as explained in Sect. \ref{sec:inertialTensor}). On the bottom panel, the same PDFs are shown but they are divided by the one of the model with $\zeta=30$ and $\log_{10}(T_{vir})=4.69897$. Each model varying $\zeta$ and $T_{vir}$ at the same time are represented here, accumulating all patches of their 101 realisations on both panels. Besides, on the top panel, the model with $\zeta=30$ and $\log_{10}(T_{vir})=4.69897$ is also represented (in gray) with a rotation of filaments in order to have a comparison with total randomness.}
    \label{fig:21mfast_5models_cos}
\end{figure}

We first look at the probability distribution functions of the patch-filament alignment for every model of the SRH set, which are shown on Fig. \ref{fig:21mfast_5models_cos}. For each model, the cosines of all of their patches are accumulated in order to improve the statistics of the result. The top panel shows that the majority of reionization patches are rather aligned with their gas filament. Besides, there does not seem to be any great difference between the different models of this set. 
The bottom panel, showing PDFs normalised to the $\zeta=30$ and $T_{vir}=5\times10^4$ K fiducial model, lets us see the slight variations between the five models. The models are well ordered in the $\cos(\langle z_{reion},\delta\rangle)\sim 1$ regime: simulations with the smaller emitting halos have more patches aligned with the density. Conversely, models with strong emitters have more frequently perpendicular configurations between matter and radiation, albeit still dominated by aligned situations.
The gray curve in the top panel corresponds to the model with $\zeta=30$ and $T_{vir}=5\times10^4$ K, for which each filament has been rotated with a randomly chosen angle for safety checks. In that case, filaments have now random orientations, which means that their orientation compared to the radiations are also random. Indeed, one can see that the PDF of this test is rather flat, depicting the expected random behaviour.

\begin{figure}
   \centering
   \includegraphics[width=0.5\textwidth]{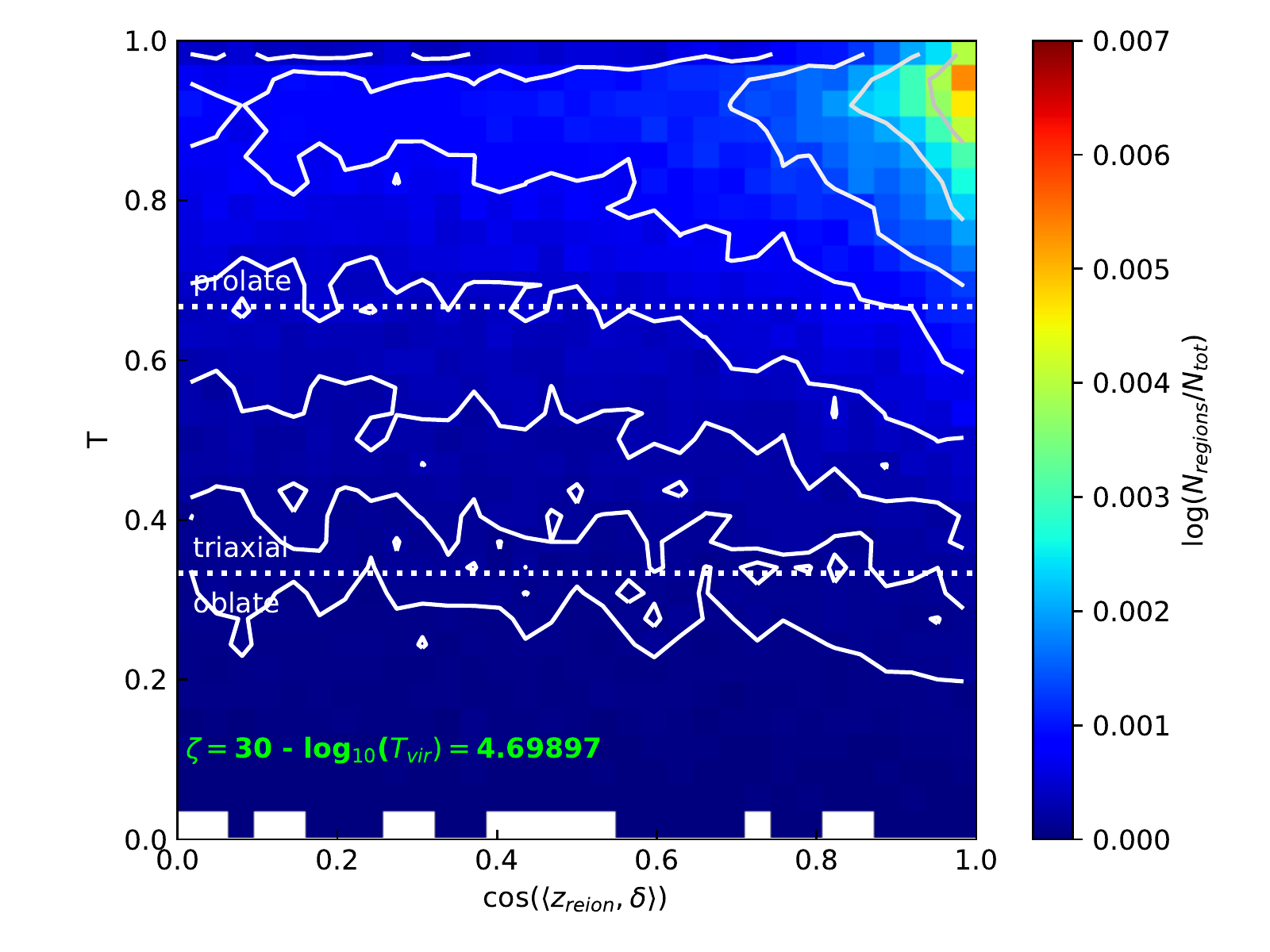}
    \caption{2D histogram representing the logarithm of the ratio of the number of patches to the total number of patches of the simulation in bins of triaxiality parameter and cosine of the angle between reionization patches and filaments for the fiducial model (the one with $\zeta=30$ and $\log_{10}(T_{vir})=4.69897$). The white full lines are the isocontours of the histogram. The coloured cells of this histogram correspond to non-zero values, and the white cells to zero values.}
    \label{fig:21cmfast_zeta30_TVsCos}
\end{figure}

\begin{figure}
   \centering
   \includegraphics[width=0.5\textwidth]{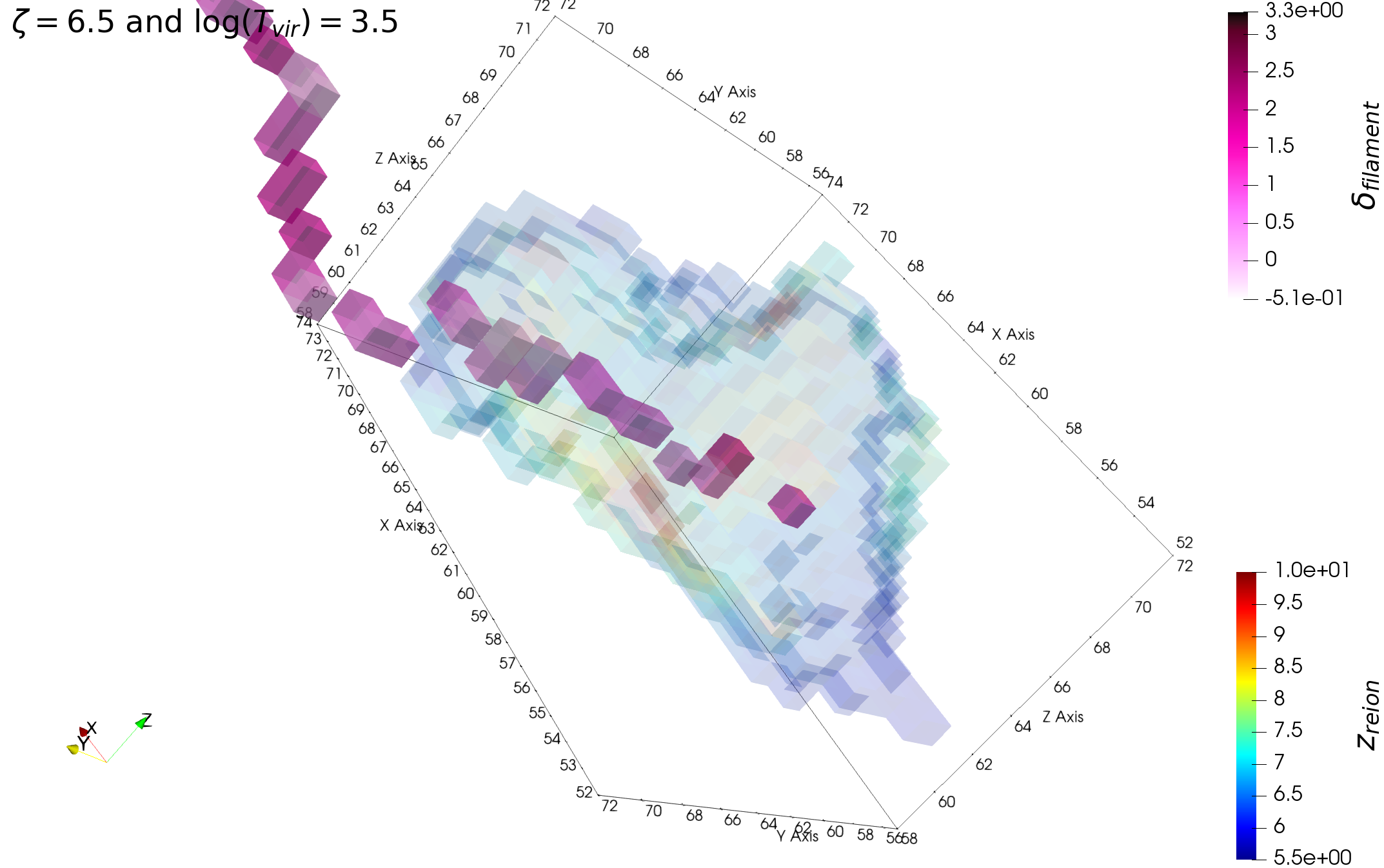}
   \includegraphics[width=0.5\textwidth]{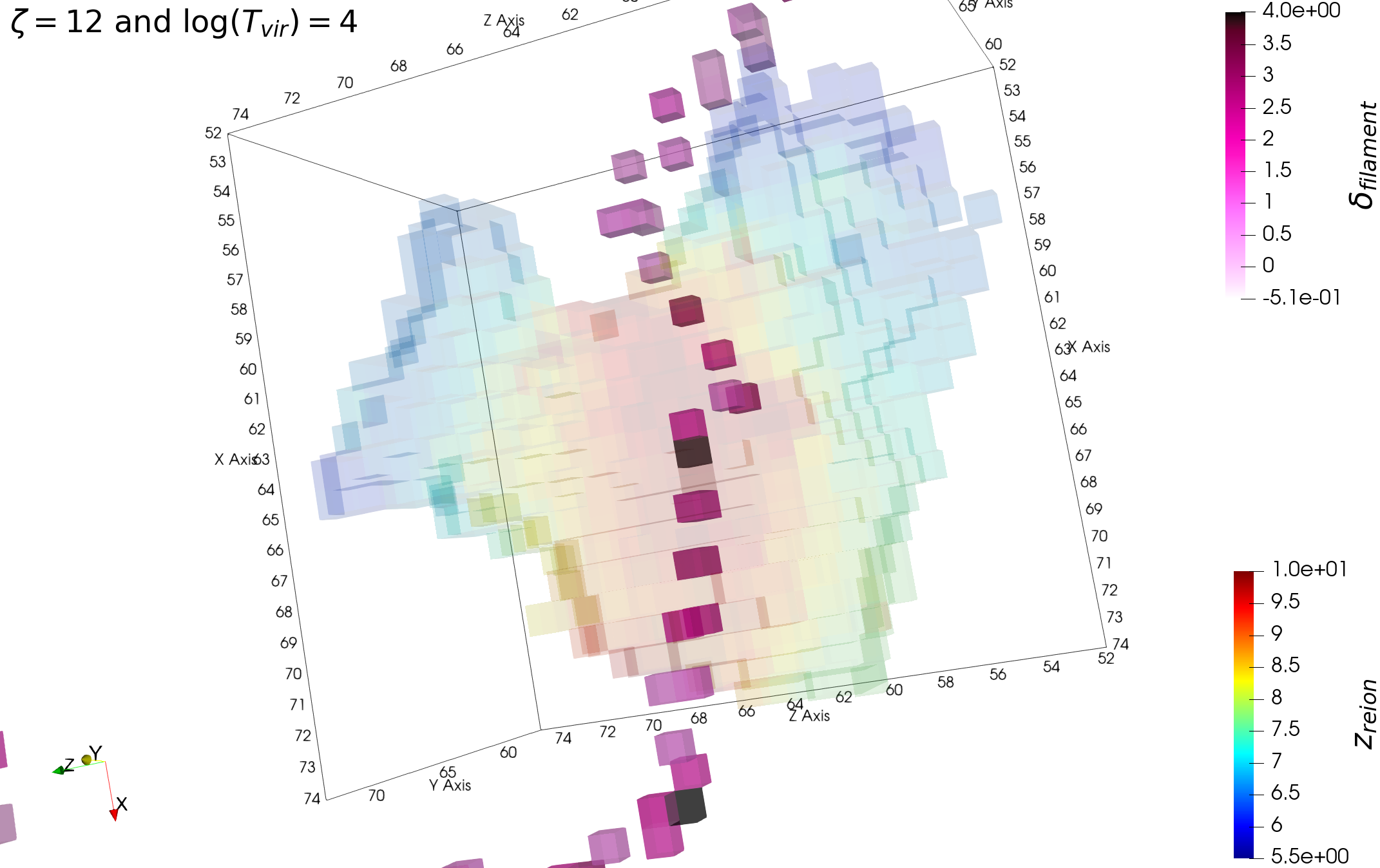}
    \caption{Examples of reionization patches and their corresponding filament extracted from the segmentations of our simulations: on the top panel, there is a prolate-aligned region, and on the bottom panel, there is an oblate-perpendicular or "butterfly" region. In shades of pink, one can see filaments, and in rainbow colors, the $z_{reion}$ region. The redder the region is, the earlier it has reionized. On the contrary, the bluer it is, the later it has reionized.}
    \label{fig:21cmfast_exCigarAndButterfly}
\end{figure}

We now know that the majority of reionization patches are aligned with their filament. In Sect. \ref{sec:inertialTensor}, we have shown that the majority of reionization patches have a prolate shape. So, is there any correlation between those behaviours? Figure \ref{fig:21cmfast_zeta30_TVsCos} is a 2D distribution of the triaxiality parameter versus the cosine of angles between the radiations and filaments for the fiducial model ($\zeta=30$ and $\log_{10}(T_{vir})=4.69897$). One can see that there is indeed a correlation between the shape and the orientation of $z_{reion}$ patches: they are generally prolate and aligned with the filaments. Note that this distribution is very similar for every other model (which are not shown here). Therefore, in our simulations there is a clear regime of prolate-aligned reionization patches (see the top-right region with $T\geq\frac{2}{3}$ and $\cos(\langle z_{reion},\delta\rangle)\geq0.5$ of the $T$ vs. $\cos(\langle z_{reion},\delta\rangle)$ space, where there is 46.6\% of the total number of patches). There are also other types of patches, even if there are rarer. For example, the opposite of the prolate-aligned patches are the oblate-perpendicular patches to the filaments (see the bottom-left region with $T\leq\frac{1}{3}$ and $\cos(\langle z_{reion},\delta\rangle)\leq0.5$ of the $T$ vs. $\cos(\langle z_{reion},\delta\rangle)$ space): there are only a few of them (2.36\%) and they are in fact patches with "butterfly" shapes. Figure \ref{fig:21cmfast_exCigarAndButterfly} shows those two kinds of patches: the pink cells belong with the filament and the rainbow-colored ones belong with the reionization patches. On the top panel, there is an example of a prolate-aligned region, where one can clearly see its prolate shape and that the reionization patch follows well the direction of the filament. On the bottom panel, the patch looks like a butterfly thanks to the two "wings" surrounding the filament: it is easier for the radiation to escape perpendicularly to the filament, as these directions are less dense. Another interesting feature with that "butterfly" patch is that one can clearly see that regions close to the filaments have been reionized earlier, while the reionization redshift decreases as the distance to the filaments increases. The same behaviour is observed on the prolate-aligned region of the top panel but in a less impressive manner. With these examples, one can clearly see that the sources are located within the filaments or very close to them and it leads to an inside-out scenario of the Reionization.

Regarding the difference between "fiber-like" and "butterfly" shapes, we believe that it originates from the population of sources within a given patch. Fiber-like regions could include a collection of radiation sources "beaded" along the filaments, with similar properties in terms of ignition time, emissivity and environment: in such situations, the source, located at the reionization peak of a given patch (i.e. the patch seed) is not exceptional compared to close-by sources within the same region. The differences in reionization times values and gradients created by these sources are not sufficient to let $\disperse$ segments a region into small individual reionization patches and it rather produces extended patches along filaments. On the other hand, butterfly patches could be the mark of a patch dominated by its seed source (because it is isolated and/or brighter than other local emitters), that singlehandedly drives the propagation of ionization fronts from the very beginning, preferentially along the path of least resistance, perpendicularly to filaments. But clearly, the latter geometry is sub-dominant in our models, a trend that can be accentuated by the late reionization scenario of the SRH dataset (compared to e.g the DRH models): at late times a greater number of locations are susceptible to host sources and patches are rarely dominated by a single source. 
It should be noted that an increase in both $T_{vir}$ and $\zeta$ can boost the propagation of ionization fronts, cancelling local small variations of reionization times and gradients around sources along filaments. It would lead to larger coherent patches and we therefore expect a decrease in the number of fibre-like regions in such cases, as measured indeed in the SRH set. 

\subsection{Different Reionization histories}
\label{sec:CosZeta}

\begin{figure}
   \centering
   \includegraphics[width=0.5\textwidth]{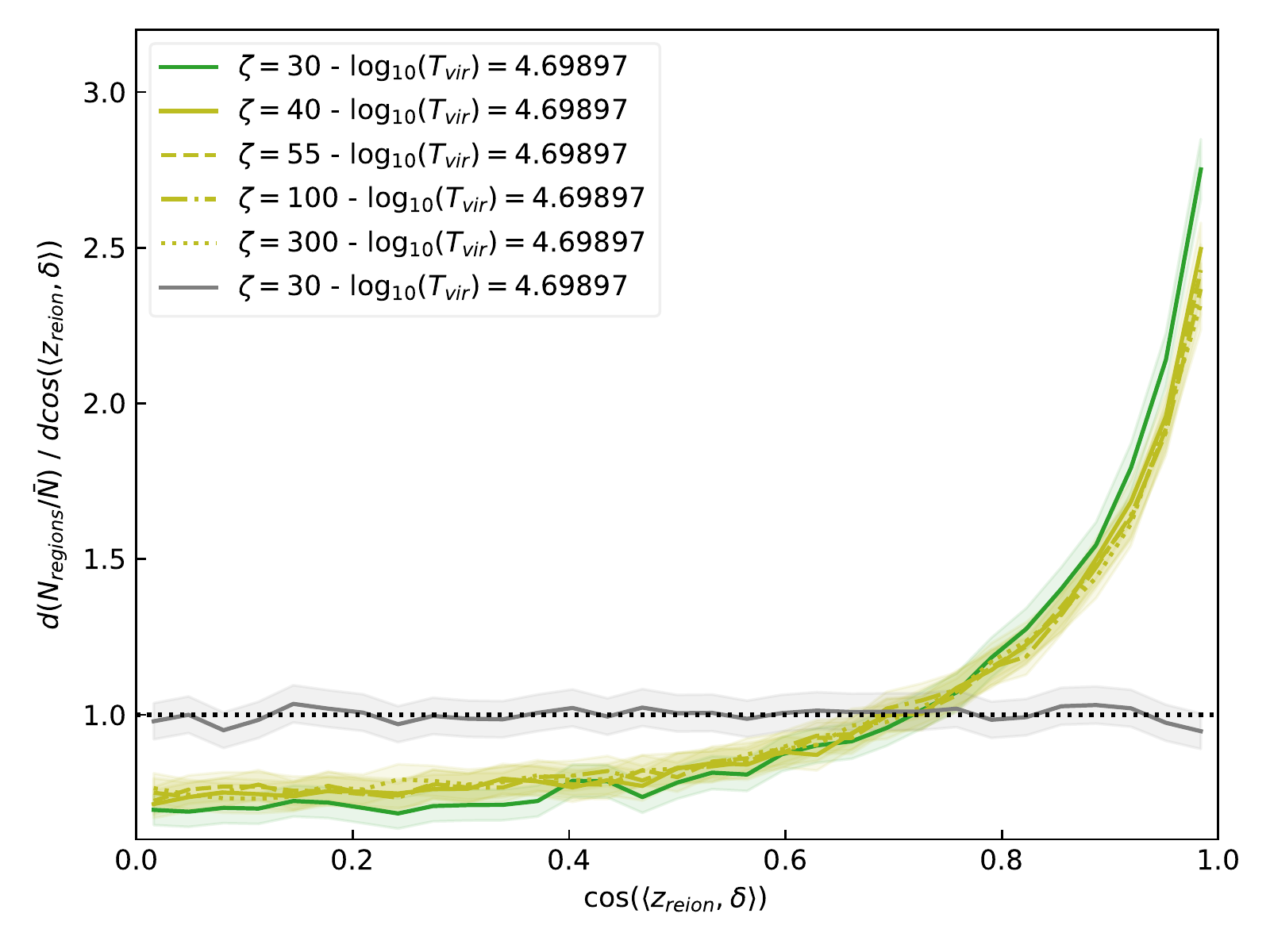}
   \includegraphics[width=0.5\textwidth]{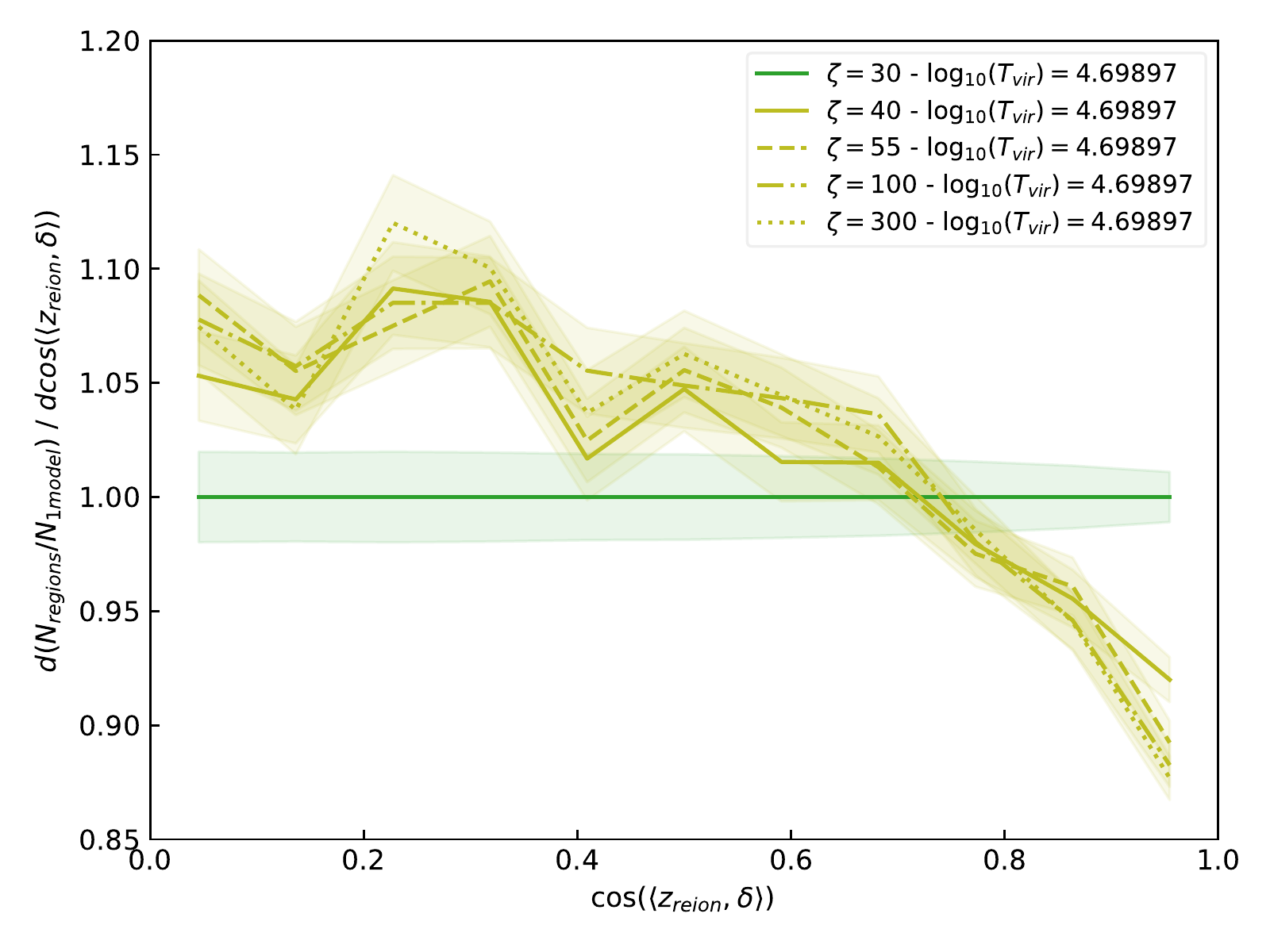}
    \caption{On the top panel, there are the probability distribution functions of the cosine of the angle between the reionization patches and their corresponding filaments (as explained in Sect. \ref{sec:inertialTensor}). On the bottom panel, the same PDFs are shown but they are divided by the one of the model with $\zeta=30$ and $\log_{10}(T_{vir})=4.69897$. Each model varying only $\zeta$ are represented here, accumulating all patches of their 101 realisations on both panels. Besides, on the top panel, the model with $\zeta=30$ and $\log_{10}(T_{vir})=4.69897$ is also represented (in gray) with a rotation of filaments in order to have a comparison with total randomness.}
    \label{fig:21cmfast_5zetas_cos}
\end{figure}

We now focus on the simulations of the DRH set (those varying only $\zeta$ for the fiducial $T_{vir}$). Changing $\zeta$ allows us to study the impact of different histories of Reionization and different halos ionizing efficiencies. 
Figure \ref{fig:21cmfast_5zetas_cos} shows the PDFs of the cosine of the angle between their reionization patches and their filaments. At first sight (top panel of this Figure), there is no significant differences between the five models, always having a majority of aligned patches with their filaments. Looking at ratios of PDFs with the fiducial model, a weak trend can be noted: the higher $\zeta$, the less the patches are aligned to the filaments and there is a slight increase in the number of perpendicular "butterfly" patches. It would be consistent with the above hypothesis regarding the origins of the different geometries: DRH models with large $\zeta$ (while keeping $T_{vir}$ constant) attribute strong emissivities to the first and rare sources that seed the patches, and let them extend their radiative influence perpendicularly to the filaments at early times.

\section{The impact of the persistence parameter}
\label{sec:persistence}

In the previously mentioned results, all the segmentations of the density and the reionization redshift fields have been obtained using $\disperse$ with a fixed value of the persistence threshold. In this section, we propose a brief comparison of the major results shown earlier but with different persistence values. This comparison is only done on the fiducial $\cmfast$ model ($\zeta=30$ and $T_{vir}=5\times10^4$ K) for the sake of brevity, but it is very likely that the conclusions remain the same for every other model.

\begin{figure}
   \centering
   \includegraphics[width=0.5\textwidth]{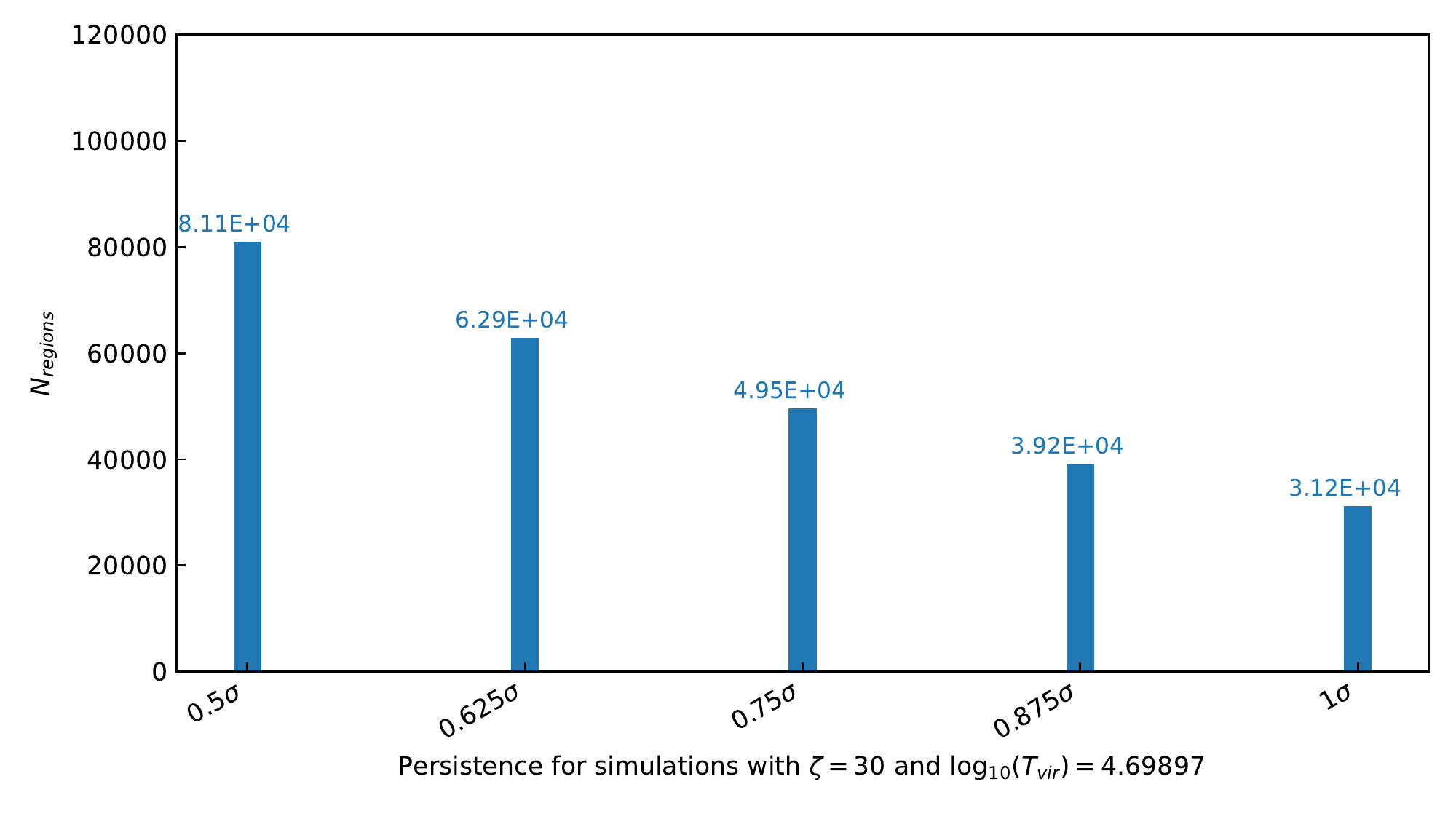}
    \caption{Number of patches in the fiducial model ($\zeta=30$ and $T_{vir}=5\times 10^4$ K) that are detected using $\disperse$ with different persistence levels. For each one of them, one adds the patches numbers of every realisation.}
    \label{fig:21cmfast_5pers}
\end{figure}

The segmentations of the density and redshift of reionization fields (of the fiducial $\cmfast$ model) are therefore computed with $\disperse$ using four other persistence values, which are 0.625$-\sigma$, 0.75$-\sigma$, 0.875$-\sigma$ and 1$-\sigma$. Figure \ref{fig:21cmfast_5pers} shows the accumulated numbers of reionization patches found in those fields for every one of the 101 realisations. One can see that the larger the persistence, the less there are patches as expected since critical points are more easily merged. It means that in the resulting segmentations with those larger persistences, there is a lower number of critical points, explaining the lower number of patches. This behaviour directly impacts the volume of the reionization patches for which the probability distribution functions are shown on Fig. \ref{fig:21cmfast_5pers_VregSizeFil}. 
The correlation of the number of patches and their volume is clearly visible here: the larger the persistence is, the larger the volume of patches is and the fewer the patches should be since the total volume is conserved.

\begin{figure}
   \centering
   \includegraphics[width=0.5\textwidth]{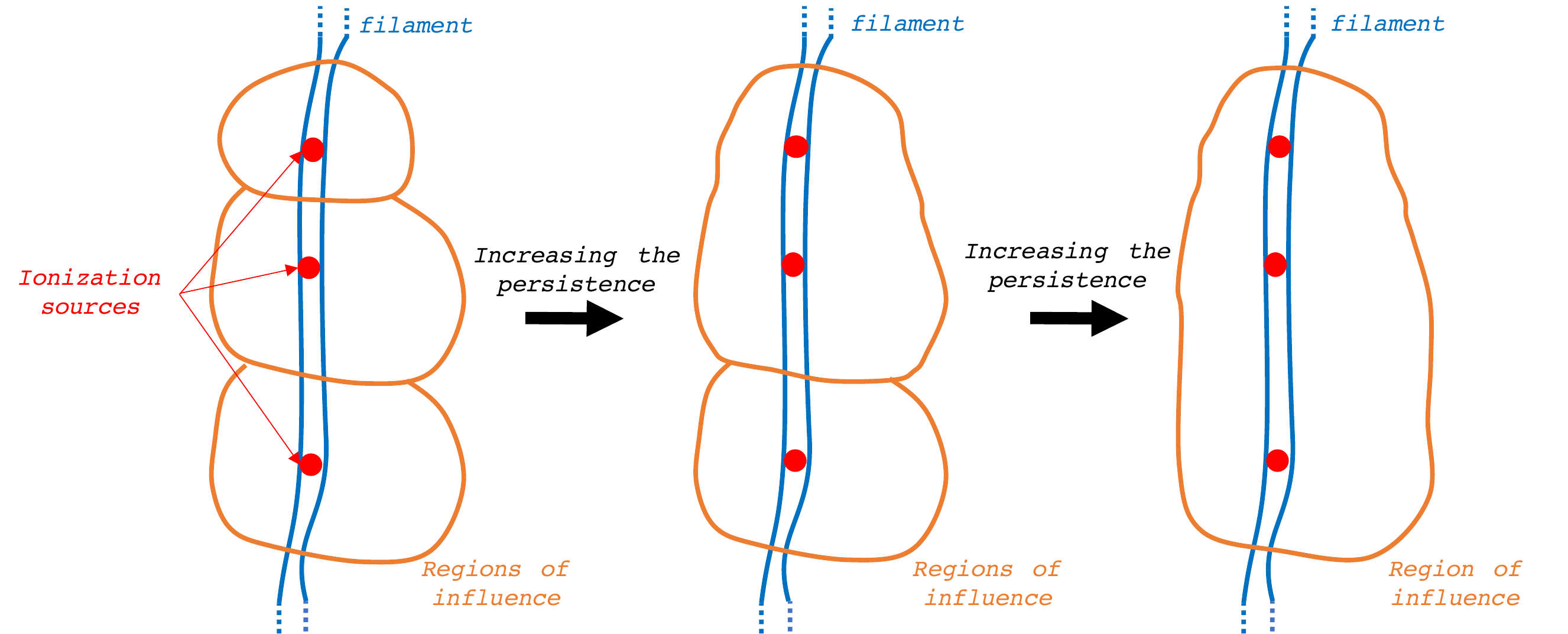}
    \caption{2D illustration of the percolation process using the persistences. A matter filament is represented in blue with many ionization seeds (in red) within it. Those seeds create their patch of radiative influence (shown in orange), and when the persistence increases (from left to right), smaller patches merge to form bigger ones.}
    \label{fig:persistence_percolation_illustration}
\end{figure}

\begin{figure}
   \centering
   \includegraphics[width=0.5\textwidth]{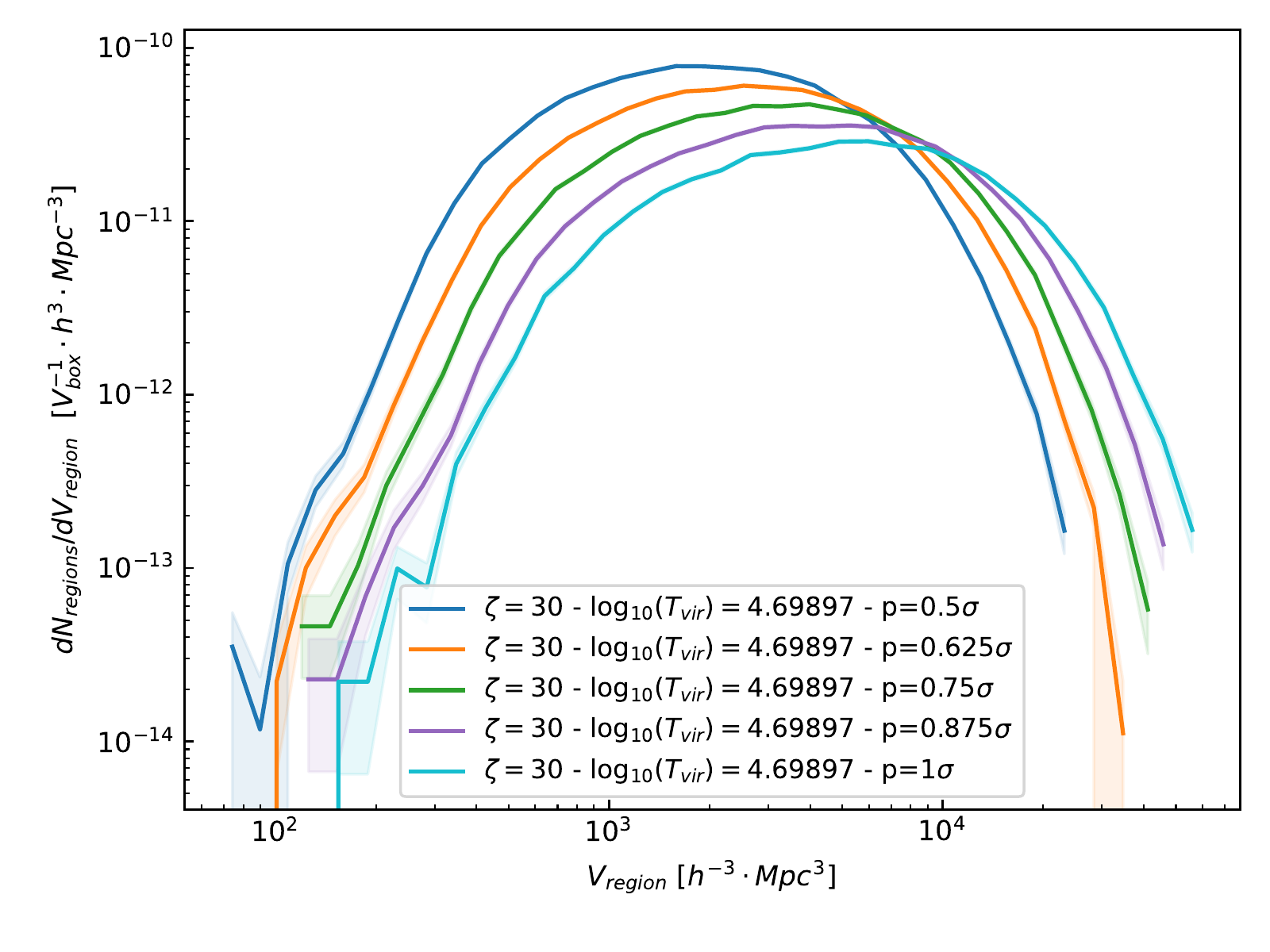}
    \caption{Probability distribution functions of the volume of segmentations patches for the fiducial model ($\zeta=30$ and $T_{vir}=5\times 10^4$ K) with the different persistences used in $\disperse$. Here, all the patches of each realisation for one specific model are accumulated.}
    \label{fig:21cmfast_5pers_VregSizeFil}
\end{figure}

It should be noted that persistence is related to percolation.
For instance, on the 1D representation of the persistence of Fig. \ref{fig:persistence_illustration}, we see that, after applying a persistence threshold, the leftmost maximum (reionization seed in our case) "disappears" and will be associated to the other one. 
Small persistence applied on the reionization redshift maps allow less prominent and more numerous patches to be taken into account in the final segmentations. Conversely, large persistence values favour the merging of patches by $\disperse$, as sketched in Fig. \ref{fig:persistence_percolation_illustration}.
The persistence applied on the reionization redshift maps can thus be seen as a relative measure of the different timings of creation of two patches and it can therefore follow the percolation process. 
We could in principle retrieve the relative progression of the fusion of ionized bubbles (regardless of the time), giving us a complementary view of the time progression of the percolation that is the subject of many studies in the literature (see e.g. \citet{Chen2019,Giri2019a,Gorce2019,Giri2020}). Indeed, \citet{Elbers2019} point out that the persistence can follow the lifetime of a topological feature and then the creation and evolution of structures, that is the percolation process. They have shown with simple models that they can follow the stages of Reionization and know in which stage the bubbles are by coupling the persistence homology with the Betti numbers.

\begin{figure}
   \centering
   \includegraphics[width=0.5\textwidth]{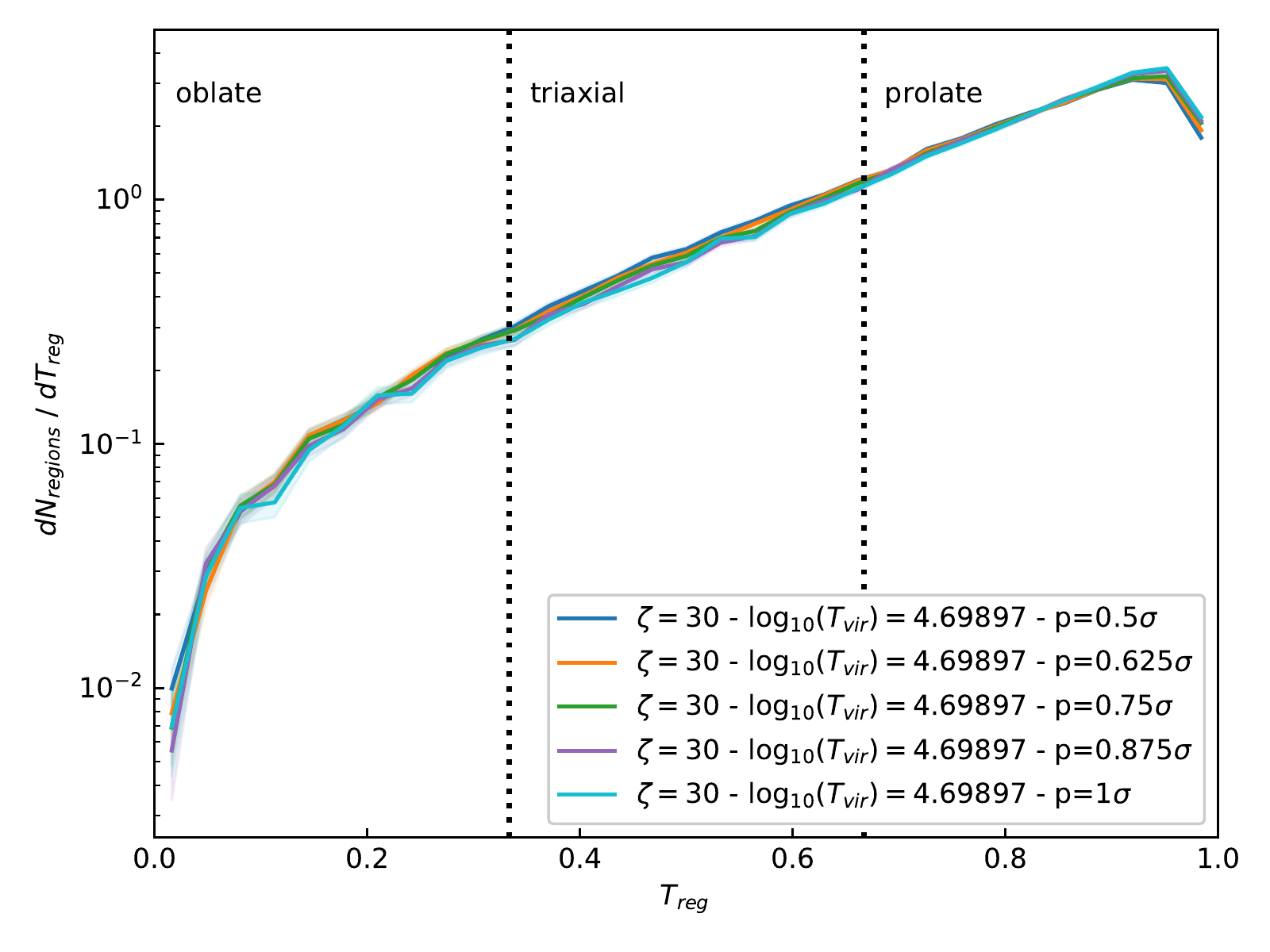}
   \includegraphics[width=0.5\textwidth]{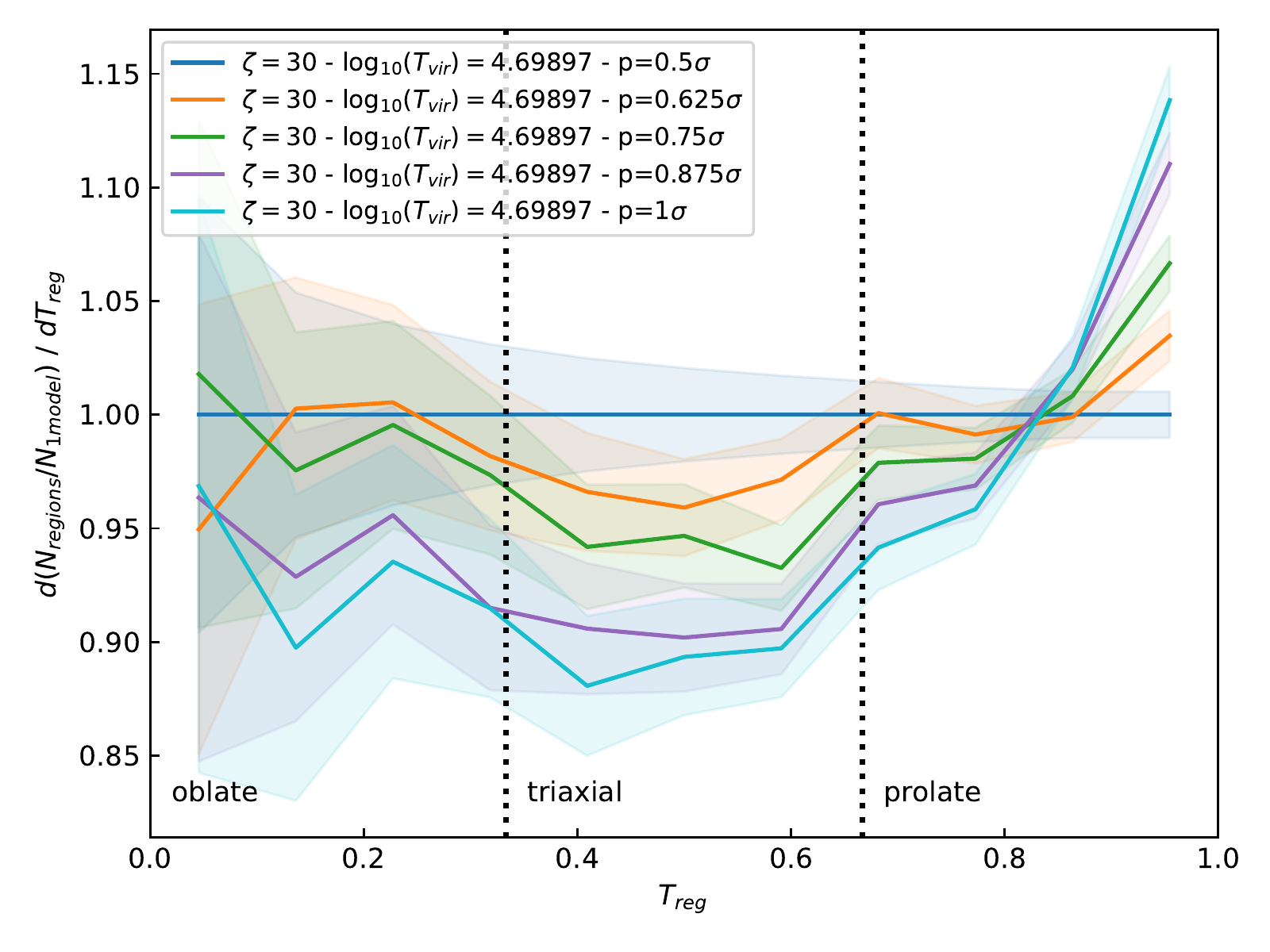}
    \caption{Probability distribution functions of the triaxiality parameter for the fiducial model ($\zeta=30$ and $T_{vir}=5\times 10^4$ K) with the different persistences used in $\disperse$ (top panel). The bottom pannel represents the same PDFs but each one of them is divided by one particular model (the fiducial one with the persistence being $0.5-\sigma$).  On the two panels, all the patches of each realisation for one specific model are accumulated.}
    \label{fig:21cmfast_5pers_T1}
\end{figure}

Now, computing the triaxiality parameter for every new persistence, one can plot the PDFs shown on the top panel of Fig. \ref{fig:21cmfast_5pers_T1}. Generally, all the segmentations with the different persistences have the same behaviour: a majority of prolate reionization patches. But looking at the detailed ratios with the 0.5$-\sigma$ curves (bottom panel of the same Figure), one can distinguish a trend with increasing persistences: the larger the persistence, the more there are prolate patches. The percolation process therefore favours the merging of patches along filaments and promotes prolate-aligned configurations, with "beaded" patches along filaments, as sketched again in Fig. \ref{fig:persistence_percolation_illustration}.
Again, our study of the persistence is reminiscent of the process detailed by \citet{Chen2019} where small ionized bubbles grow during their "Ionized Bubble" stage and merge to form the bigger and fibre-like regions during the "Ionized Fibre" stage. Changing the persistence parameter allows us to retrieve a similar behaviour in our segmentations.


\begin{figure}
   \centering
   \includegraphics[width=0.5\textwidth]{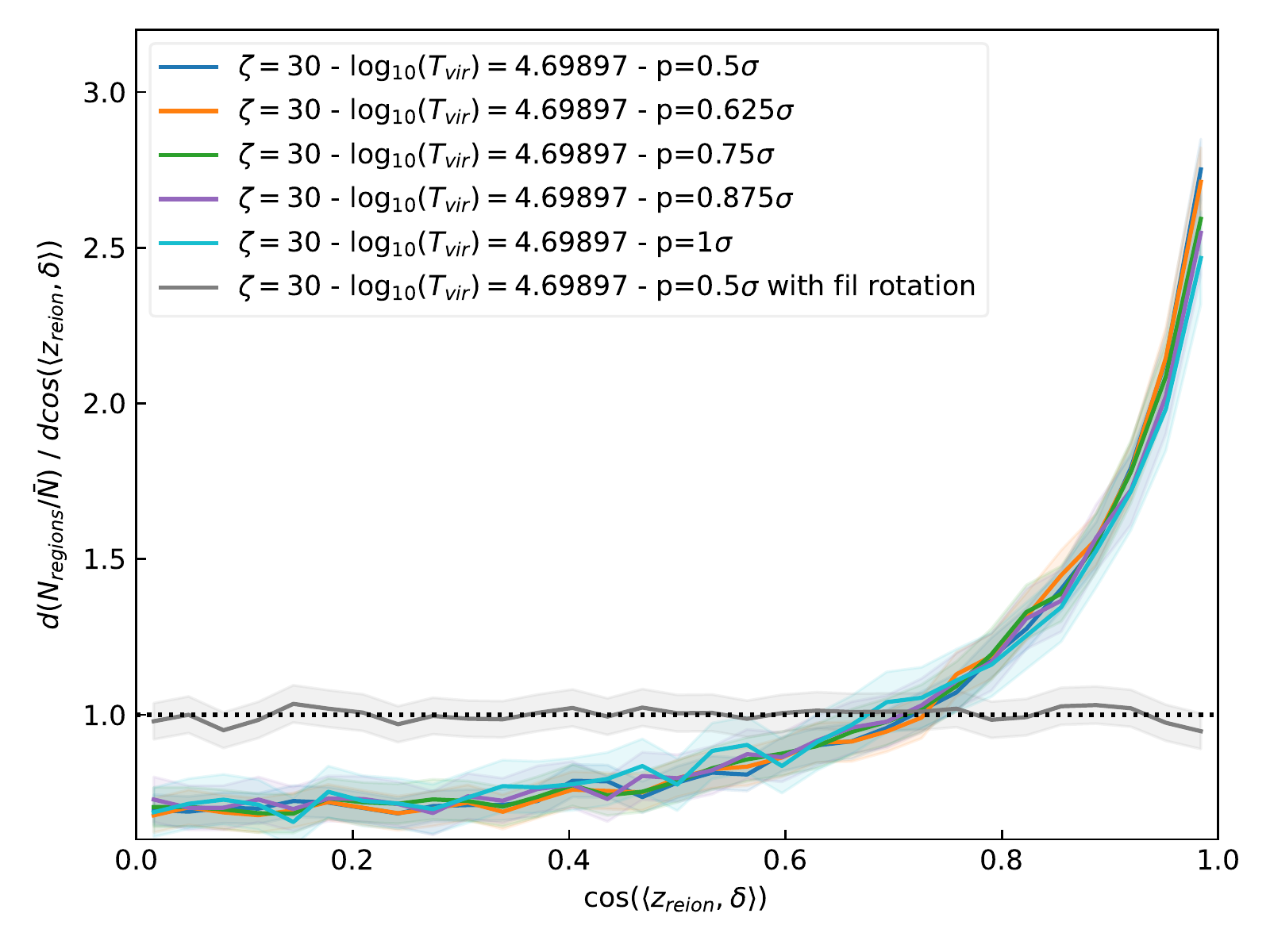}
    \caption{Probability distribution functions of the cosine of the angle between the reionization patches and their corresponding filaments (as explained in Sect. \ref{sec:inertialTensor}) of the fiducial model ($\zeta=30$ and $T_{vir}=5\times 10^4$ K) with the different persistences used in $\disperse$, accumulating all patches of every realisation. Besides, the model with the $0.5-\sigma$ is also represented (in gray) with a rotation of filaments in order to have a comparison with total randomness.}
    \label{fig:21cmfast_5pers_cos}
\end{figure}

Nevertheless, our segmentations appear to be robust to the changes in persistence, and the main conclusions that have been mentioned in previous sections. are unchanged. Indeed, the shape of the reionization patches (see Fig. \ref{fig:21cmfast_5pers_T1}) remains very similar and the orientation of the patches with respect to the filaments is almost unchanged too, as one can see on Fig. \ref{fig:21cmfast_5pers_cos}.

%--------------------------------------------------------------------
%--------------------------------------------------------------------
%--------------------------------------------------------------------
\section{Comparison with $\emma$ simulations}
\label{sec:EMMAsimus}

We now compare the topology of EoR simulations obtained with the semi-numerical code $\cmfast$ to that coming from the cosmological simulation code $\emma$ (Electromagnétisme et Mécanique sur Maille Adaptative, \citet{Aubert2015}). $\emma$ is an adaptive mesh refinement (AMR) code that couples radiative transfer and hydrodynamics, and in which the light is described as a fluid (resolved using the moment-based M1 approximation, \citet{Aubert2006}).

The $\emma$ simulations set on which we work here is composed of two simulations with cosmology given by \citet{PlanckCollaboration2020}, with no AMR and no reduced speed of light (see \citet{Gillet2021} for more information on the simulations). The difference between the two simulations is the mass resolution of the stellar particles: $10^{7}$ M$_\odot$ for the one labelled as $\emma$ "Mslow"  and $10^{8}$ M$_\odot$ for the other one. 
As detailed in \citet{Gillet2021}, stellar particles are converted from gas according to a Poissonian process. Cells that will convert to stars an amount of gas only slightly larger than the stellar mass will naturally lead to a stochastic star formation process. Conversely, if the same amount of gas is much larger than the stellar mass particle, it will induce a steady stream of particles. As a consequence the $\emma$ "Mslow" model presents a diffuse population of stars, whereas the other one exhibits a stellar biased population focused on the densest cells.
For these two simulations, the boxes have 512$^3$ cells for 512$^3$ cMpc$^3$/h$^3$. We have then cut those boxes in 64 sub-boxes of  128$^3$ cells and 128$^3$ cMpc$^3$/h$^3$, creating 64 "realisations" of the two $\emma$ models. The resolution of those sub-boxes is the same than the one of the $\cmfast$ simulations. Comparisons with the $\cmfast$ models are made with 64 realisations of each $\cmfast$ simulation for consistency.

\begin{figure}
   \centering
   \includegraphics[width=0.5\textwidth]{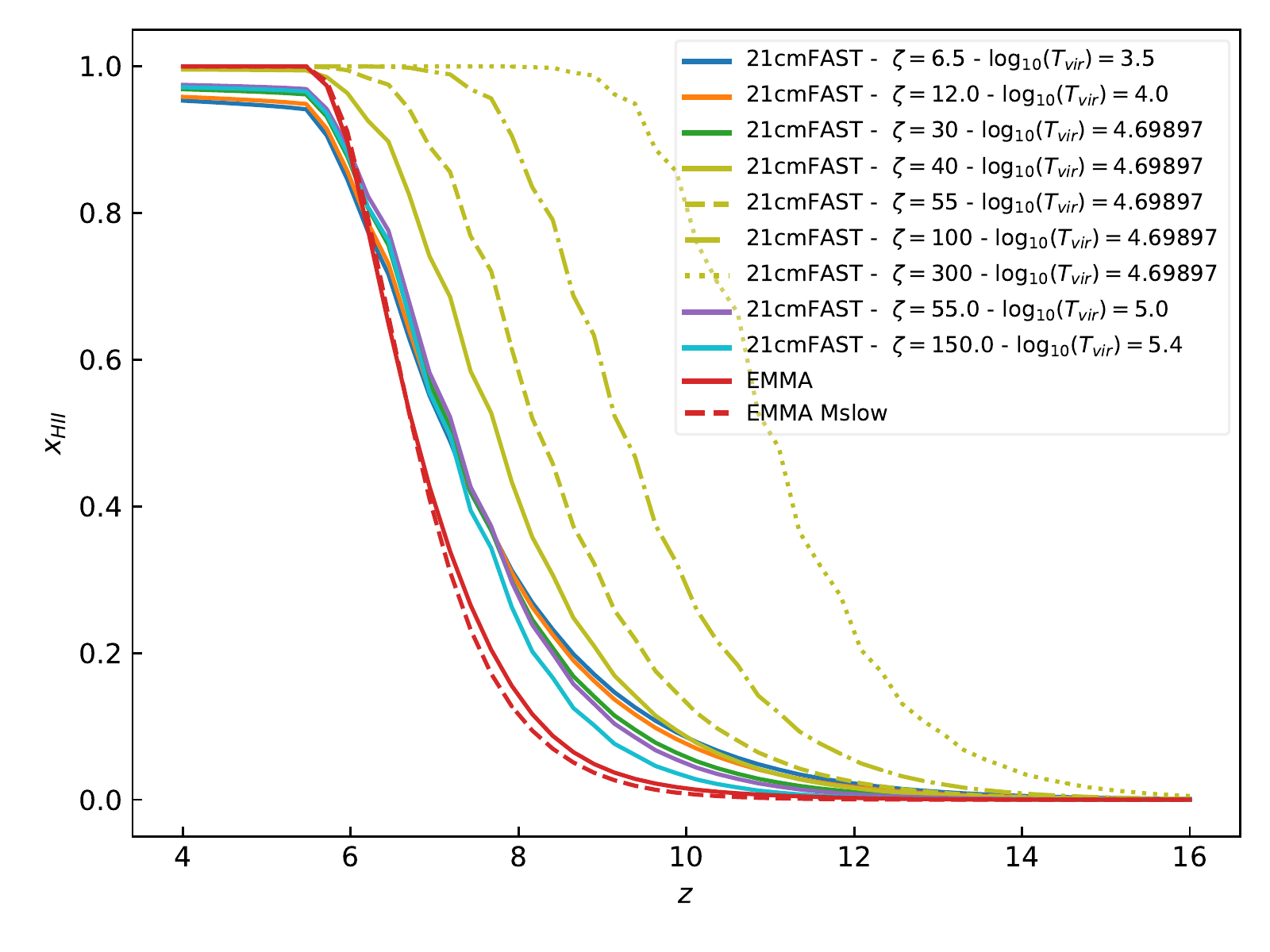}
    \caption{Ionized volume fraction of all the $\cmfast$ simulations and the two EMMA simulations having different models, each one of them averaged on 64 realisations.}
    \label{fig:EMMA_xion}
\end{figure}

On Fig. \ref{fig:EMMA_xion}, one can see the ionized volume fractions of every simulation mentioned in this study. The $\emma$ ones are fully ionized approximately at the same time as the $\cmfast$ ones. With $\emma$, however, the reionization starts later and is more abrupt, which could be explained by the fact that $\emma$ simulations have larger volumes compared to the $\cmfast$ ones, with greater voids and larger halos.

\begin{figure}
   \centering
   \includegraphics[width=0.5\textwidth]{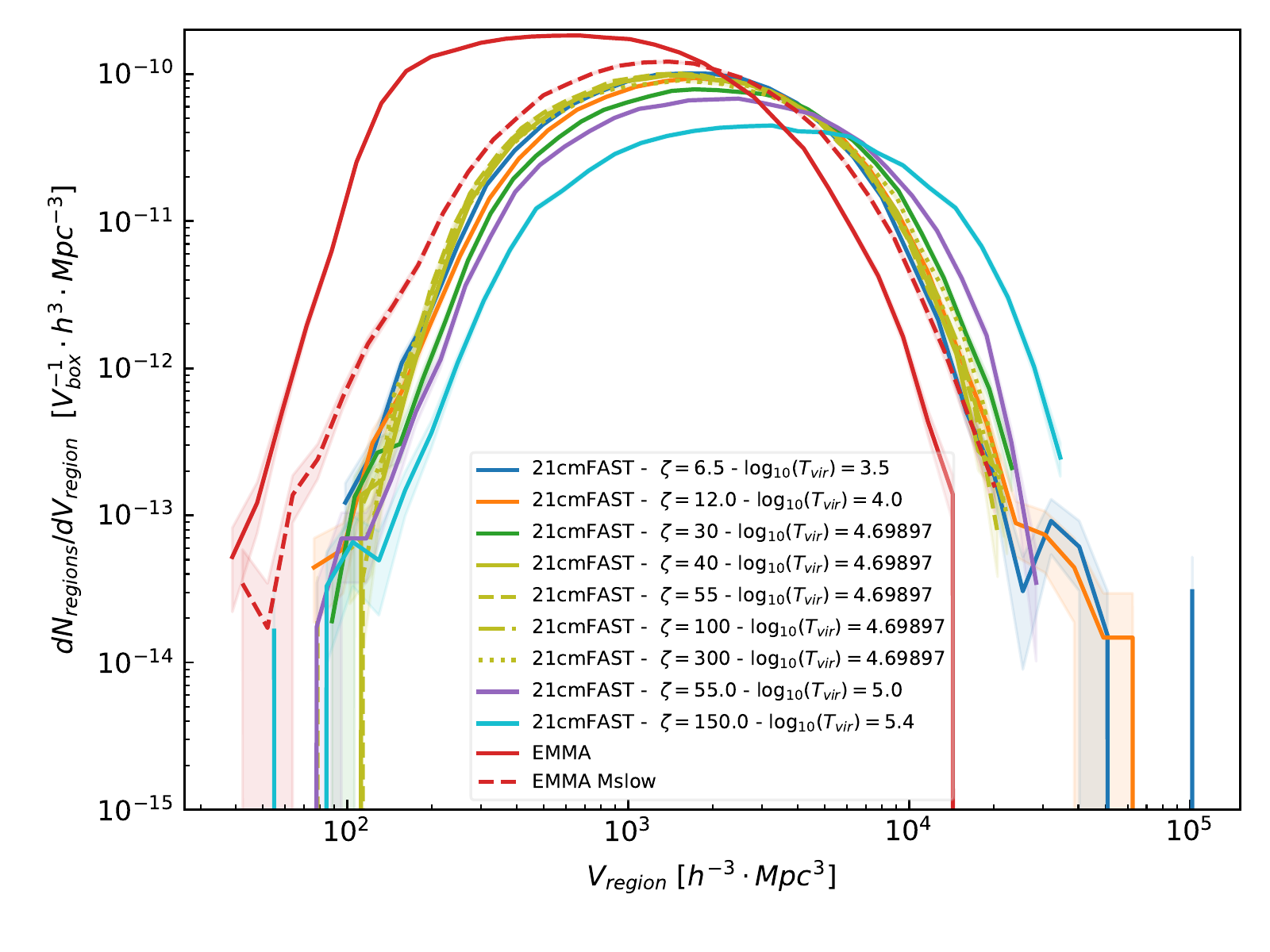}
    \caption{Probability distribution functions of the volume of segmentations patches for every $\cmfast$ and $\emma$ simulation. Here, all the patches of each 64 realisations for one specific model are accumulated.}
    \label{fig:EMMA_VregSizeFil}
\end{figure}

Again, before using $\disperse$ on the $\emma$ fields, the same kind of transformations as mentioned in Sect. \ref{sec:disperse} are applied on each sub-box, using the means and the standard deviations of the 512$^3$ cMpc$^3$/h$^3$ boxes instead of those of each sub-grid. Doing so, we take into account the cosmic variance of the whole box.
After having obtained the segmentations with $\disperse$ with the same persistence used for all the $\cmfast$ simulations (0.5$-\sigma$), one can compute the distribution of volumes of patches. It is shown in Fig. \ref{fig:EMMA_VregSizeFil} for every $\emma$ and $\cmfast$ simulations accumulating 64 realisations for each one of them. The $\emma$ simulations have smaller patches and a greater number of them than the $\cmfast$ simulations. The $\emma$ model with the massive, $10^8 M_\odot$, stellar particles is significantly different from the $\cmfast$ simulations: the patch size distribution is clearly shifted to smaller volumes with a large number of small patches. Meanwhile, the $\emma$ "Mslow" model with the less massive, $10^7 M_\odot$, stellar particles is more alike the $\cmfast$ models: the more diffuse source distribution, allowed by such settings, manage to recover topological features similar to $\cmfast$ models (in which all locations is a potential source of radiation with its own distribution of emitting halos). 

\begin{figure}
   \centering
   \includegraphics[width=0.5\textwidth]{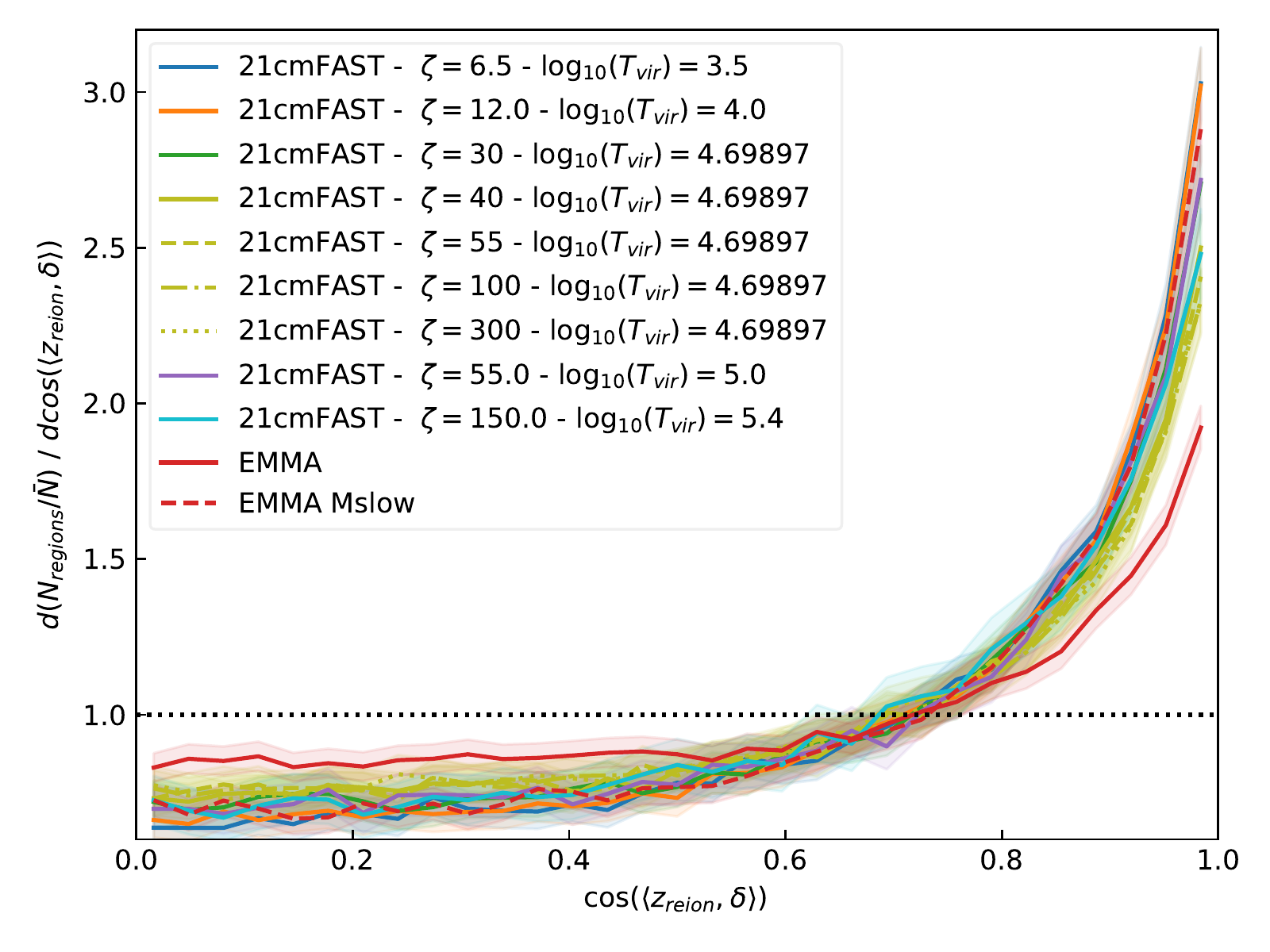}
   \includegraphics[width=0.5\textwidth]{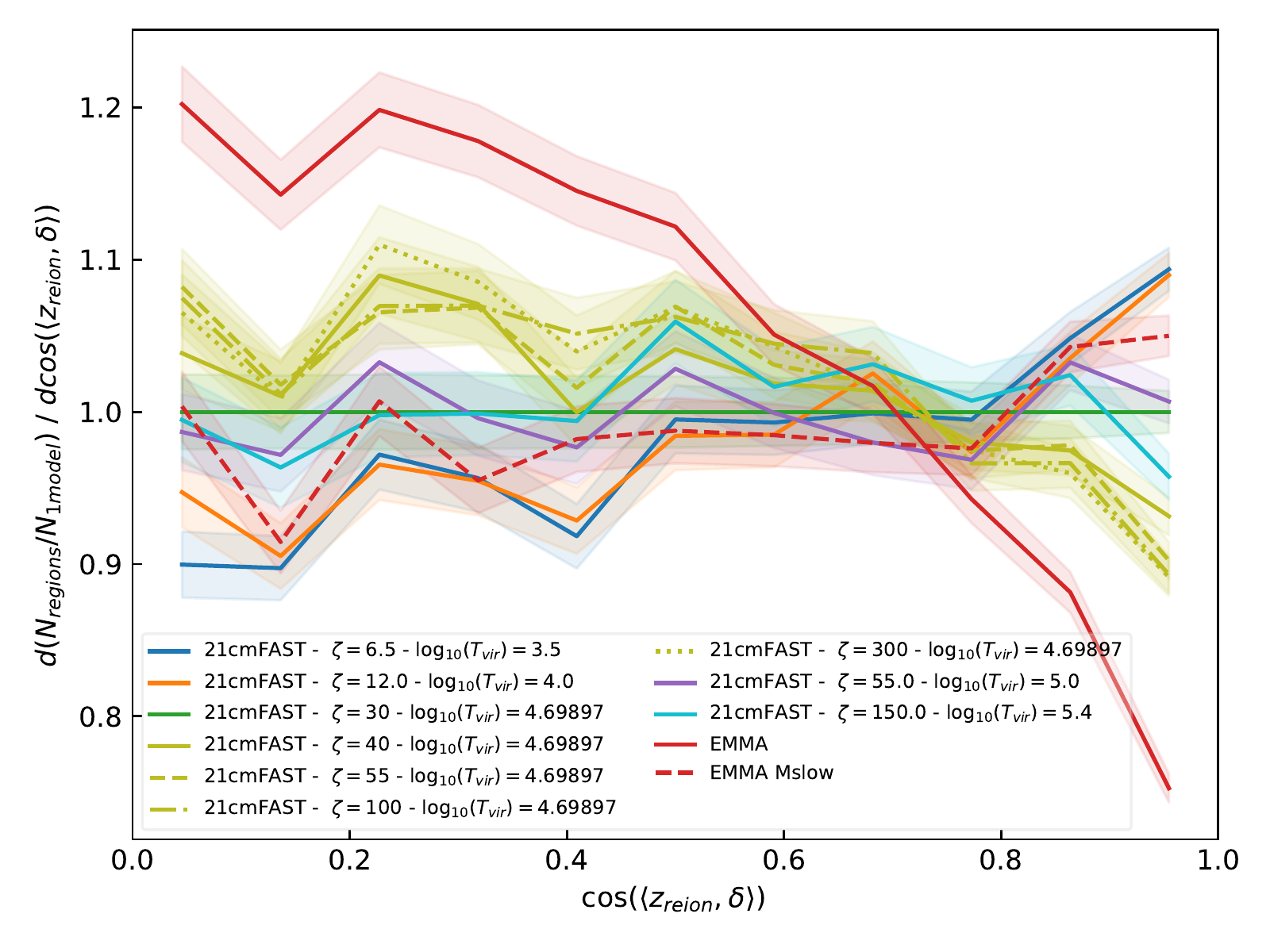}
    \caption{On the top panel, there are the probability distribution functions of the cosine of the angle between the reionization patches and their corresponding filaments (as explained in Sect. \ref{sec:inertialTensor}). On the bottom panel, the same PDFs are shown but they are divided by the one of the model with $\zeta=30$ and $\log_{10}(T_{vir})=4.69897$. Each $\cmfast$ model is represented here, alongside with the two $\emma$ simulations, accumulating all patches of 64 realisations on both panels.}
    \label{fig:EMMA_cos}
\end{figure}

The tensors and main directions of the reionization patches and filaments of the $\emma$ simulations are computed, and allow us to compare the direction of the propagation of radiation with respect to the filament. The top panel of Fig. \ref{fig:EMMA_cos} shows the distributions of the cosine of the angle between those directions for the $\emma$ and the $\cmfast$ simulations. There is an alignment of patches and gas filaments in the $\emma$ simulation similar to the $\cmfast$ models, with the $\emma$ "Mslow" being slightly more aligned. The bottom panel of Fig. \ref{fig:EMMA_cos} shows the ratio of each PDFs of cosines with the fiducial $\cmfast$ model and again the $\emma$ "Mslow" shows very similar PDFs to the $\cmfast$ models, which implies that it is possible to recover the same kind of behaviour than semi-analytical simulations with a fully-numerical code even if the reionization occurs differently. 
Conversely, the $\emma$ model with massive, $10^8 M_\odot$, stellar particle presents quite a different picture in terms of alignments. The scarce, stochastic and biased distribution of sources in this models leads to much more perpendicular patches than all the other models. While this model is in reasonable agreement in terms of reionization history, its topology ends up being quite different: it emphasises how models that present some reasonable global agreements can end up being vastly different in terms of topology as soon as the details of for example the source models are not equivalent.
The higher star formation stochasticity and the more biased distribution of sources lead to trends similar to the larger virial temperatures in the $\cmfast$ simulations, with isolated reionization seeds that dominate more their environment.
As for the shape of the reionization patches in the $\emma$ simulations set, they are also rather prolate in both simulations as in the $\cmfast$ simulations and no significant difference between the models have been found (not shown here).

%--------------------------------------------------------------------
%--------------------------------------------------------------------
%--------------------------------------------------------------------
\section{Conclusions}
\label{sec:conclusion}

In this study, we have analysed the topological structures extracted from different EoR models made with the $\cmfast$ semi-analytical code and the $\emma$ cosmological simulation code. For that purpose, we have used the reionization redshift and gas density fields. The reionization redshift field corresponds to the time when the gas has reionized, and contains spatial and temporal information on the physical processes during the EoR. To extract the topological information, we have used $\disperse$, that applies concepts of the Morse theory, on the density field to get the filaments of matter and on the reionization redshift field to segment it into reionization patches coming from the same seed of local reionization.

In the $\cmfast$ models, the size distributions of the reionization patches show us that they have an average radius of 10 cMpc/h quantifying the typical extent of the radiative influence of the seeds of reionization. In general increasing the virial temperature and the ionizing efficiency of halos decreases the number of patches and makes them larger, extending the reach of the radiative influence of the seeds of local reionization. 

Using the inertial tensors of the reionization patches and density filaments, one could extract their shape and orientation. We found that the reionization patches are statistically prolate and aligned with the local gas filaments. Radiation escapes from sources beaded along the gas filament to reach the zones that reionize the latest (typically, the voids) in an inside-out manner. A minority of patches have the opposite behaviour with a "butterfly" shape, that are oblate patches with radiations escaping perpendicularly to the local filaments. We believe that the difference arises from the properties of sources within the patches: if the initial source that defines a reionization patch is locally dominant in terms of photons output, butterflies shape can appear, whereas if the initial source is only the first of a collection of rather similar close-by emitters, patches end up aligned with the local distribution of matter. We note that the dominant status of a source can depend on many factors (intrinsic emission, escape fraction, density  and configuration of local absorbers) and is therefore likely to be resolution and/or model dependent. Our results, obtained at rather low resolution ($1.48^3$ cMpc\textsuperscript{3}), should be revisited later on, for instance at higher resolution.

Comparisons of the $\cmfast$ segmentations with the ones obtained from $\emma$ simulations showed that $\emma$ simulations can provide results similar to the semi-analytical predictions in terms of patches properties (volume, number) and relative orientations to the filaments of matter. We also find that $\emma$ simulations with increased bias and stochasticity for star formation can also lead to quite different topologies on the same criteria, even though they present similar reionization histories.

These results appear to be robust when changing the persistence parameter with $\disperse$. As the persistence increases and patches are merged, the latter gets larger and more prolate. It provides an insight on the percolation process of our EoR models, directly from a single reionization map. We find that during the first stages of the Reionization, isolated ionized bubbles appear and merge with time in order to form larger and fibre-like ionized bubbles \citep{Chen2019}. This subject could be further investigated, as it is complementary to standard percolation studies based on sequences of snapshots of ionized fractions maps.

To sum up, the reionization redshift field contains a lot of spatial and temporal information, and is found to be quite useful in the current topological framework. In addition, making statistical analyses with this field is quite practical since it requires to only extract one $z_{reion}$ field from each simulation. We believe that we only skimmed through the potential of the approach developed here.
We already mention the possibilities offered by the persistence parameter. We could also investigate the halo properties in relation to their host reionization patches. 
Moreover, we could work on smaller scales and see if the study of the patches geometry (size, shape, orientation) could inform us about the ways radiations escape from galaxies, how far they can influence their environment and even about the time evolution/duty cycles of the photons production (see e.g. \citet{Deparis2019}). However, such studies on galaxies or halos are not possible at the resolution discussed here  and ideally would be conducted with models with CoDa-like characteristics \citep{Ocvirk2016}.
Finally the $z_{reion}$ field is not a direct observable, and similar analyses could be directly done on the topology of 21-cm maps or lightcones. These geometrical studies may reveal a way to relate these observations to the timing of the reionization process.

\section*{Acknowledgements}
We thank Adélie Gorce for her proofreading and her precious questions and advice.
We also thank Christophe Pichon for his help with the topological code $\disperse$.
Furthermore, we express our gratitude to the referee of this paper for the useful and constructive comments that allowed us to make a more comprehensive and complete paper.
This work was granted access to the HPC resources of CINES under the allocations 2020-A0070411049 and 2021- A0090411049 “Simulation des signaux et processus de l’aube cosmique et Réionisation de l’Univers” made by GENCI.
This research made use of \textsc{astropy}, a community-developed core Python package for astronomy \citep{AstropyCollab2018}; \textsc{matplotlib}, a Python library for publication quality graphics \citep{Hunter2007}; \textsc{scipy}, a Pythonbased ecosystem of open-source software for mathematics, science, and engineering \citep{Virtanen2020}; \textsc{numpy} \citep{Harris2020} and \textsc{Ipython} \citep{Perez2007}.
   
% WARNING
%-------------------------------------------------------------------
% Please note that we have included the references to the file aa.dem in
% order to compile it, but we ask you to:
%
% - use BibTeX with the regular commands:
\bibliographystyle{aa} % style aa.bst
\bibliography{biblio.bib} % your references Yourfile.bib
%
% - join the .bib files when you upload your source files
%-------------------------------------------------------------------

%--------------------------------------------------------------------
%--------------------------------------------------------------------
%--------------------------------------------------------------------
\end{document}